\documentclass[aps,prb,twocolumn,showpacs,preprintnumbers,prb,superscriptaddress]{revtex4-2}
\usepackage{graphicx}  
\usepackage{dcolumn}   
\usepackage{bm}        
\usepackage{amssymb}   
\usepackage{framed}
\usepackage{amsmath}
\usepackage{hhline}
\usepackage{xcolor}
\definecolor{mygreen}{RGB}{60,150,60}
\usepackage{amsmath,verbatim,moreverb}
\usepackage{tabularx}
\usepackage{lipsum}
\usepackage{threeparttable}
\usepackage{booktabs,amssymb,siunitx}
\usepackage{latexsym}
\usepackage{dcolumn}
\usepackage{subfig}
\usepackage{epsf}
\usepackage{float}
\usepackage{url}
\usepackage{hyperref}
\usepackage{caption}
\usepackage{xcolor}
\usepackage{tikz}
\usepackage[utf8]{inputenc}
\usepackage[normalem]{ulem}
\usepackage{ragged2e}
\usepackage{etoolbox}
\newcounter{subeqn} %

\usepackage{mathptmx}
\usepackage{tikz}
\usetikzlibrary{arrows.meta, calc, decorations.pathmorphing}
\usepackage{amsmath}
\usepackage{amssymb}
\usepackage[T1]{fontenc}
\usepackage{booktabs}
\usepackage{tabularx}
\usepackage{adjustbox}

\usepackage{hyperref}
\hypersetup{
	unicode=true,          
	pdftoolbar=true,        
	pdfmenubar=true,        
	pdffitwindow=false,     
	pdfstartview={FitH},    
	colorlinks=true,       
	linkcolor=blue,          
	citecolor=blue,        
	urlcolor=blue,
}

\begin{document}

	\title{
		Multicharged Kramers--Weyl Fermions, Quantum Geometric Hall Responses and de Haas--van Alphen Oscillations in WAl$_2$
	}
	\author{Saurav Patel}
	\email{sauravpatel317@gmail.com}
	
	\author{Meet Shah}
	\email{meetshahwork2007@gmail.com}
	
	
	\author{Prafulla K. Jha}
	\email{prafullaj@yahoo.com}
	\affiliation{Department of Physics, Faculty of Science, The Maharaja Sayajirao University of Baroda, Vadodara-390002, Gujarat, India.}

\begin{abstract}

Quantum geometry of Bloch wave functions and crystalline symmetries play a central role in modern condensed matter physics, jointly shaping the electronic topology and physical responses of solids. In particular, Kramers-Weyl fermions, spin-orbit-split Fermi surfaces probed through quantum oscillations, and quantum geometric nonlinear Hall and intrinsic spin Hall responses have garnered significant attention. Despite progress in understanding these phenomena individually, their coexistence with topological surface states, Fermi arcs, and chirality dependent quantum geometric transport remains scarce, as pronounced responses involve suitable symmetries, spin-orbit coupling, and Fermi surface geometry. Here, we explore this synergy in experimentally synthesized WAl$_2$, the first aluminide with a chiral C40 structure, considering both structural enantiomers using \textit{first-principles} calculations and symmetry analysis. We observed \textit{twelve} Kramers--Weyl (KW) crossings with quantized chiral charges $|C|=1$ and $|C|=3$, accompanied by nontrivial surface states and Fermi arcs. The pronounced metallic character gives rise to \textit{six Fermi surface sheets} exhibiting antisymmetric spin--orbit coupling (ASOC), with distinct extremal orbits and candidate spin-split pairs resolved through calculated de Haas--van Alphen (dHvA) quantum oscillations. The electronic structure also exhibits large intrinsic Berry curvature dipole (BCD) arising from the asymmetric distribution of Berry curvature, giving rise to a second-order nonlinear Hall response. The opposite structural handedness of the two enantiomers reverses the topological chiralities and BCD, consequently reversing the direction of Fermi arcs and nonlinear Hall current. The BCD attains $|D_{zz}|\approx0.338$, yielding estimated nonlinear Hall current densities of $|j_x|\approx5.04\times10^{10}~\mathrm{A\,m^{-2}}$ and $|j_y|\approx2.52\times10^{10}~\mathrm{A\,m^{-2}}$. Furthermore, we observe enhanced intrinsic spin Hall conductivity (SHC) associated with the energetic alignment of the triply charged Kramers--Weyl point at $\Gamma$, suggesting a role of higher Chern number in the spin Hall response. The calculated SHC ($254.83\,(\hbar/e)\,\mathrm{S\,cm^{-1}}$) around the Fermi is largest among other C40 chiral materials. These findings underscore the broader ramifications of crystal chirality and quantum geometry, motivating further theoretical and experimental investigations into their interplay in shaping topological and transport phenomena in quantum materials.
\end{abstract}

\maketitle
	
\section{Introduction}
Topological quantum materials have established a paradigm in which the electronic properties of solids are determined not only by the energy dispersion of Bloch bands, but also by the symmetry and geometry of their quantum wave functions\cite{Gao2026QG,Liu2025QG,Palvi2023}. In momentum space, the evolution of Bloch states is encoded by quantum geometric quantities such as the Berry connection and associated Berry curvature, whose integrals define the Berry phase and over closed manifolds, topological invariants such as the Chern number \cite{Berry1984,Xiao2010,Gao2026QG}. This geometrical description has provided a unified basis for understanding a broad range of phenomena; from Dirac and Weyl quasiparticles to anomalous, spin, and nonlinear Hall responses \cite{Armitage2018,Gao2026QG,Ulrich2026}. Crystal symmetry plays a central role in this framework because it determines where band degeneracies can occur, how they are protected, and which quantum geometric responses are symmetry allowed \cite{Gao2026QG,Liu2025QG}. In particular, chiral crystals provide a natural setting in which structural handedness in real space is connected to the chirality of electronic states in momentum space\cite{Chang2018,Hasan2021}.
	
Chiral crystals lack inversion, mirror and roto-inversion symmetries and occur as enantiomorphic structures of opposite handedness. This structural handedness can manifest  in optical activity and magnetochiral responses \cite{Zhang2024Chirality}, chirality dependent coupling to circularly polarized light \cite{Pu2026}, chirality-induced spin polarization \cite{Pu2026,Kousaka2023}, nonlinear optical responses \cite{Pu2026}, and nonlinear Hall transport \cite{Joseph2024}. In nonmagnetic chiral crystals with spin orbit coupling (SOC), Kramers degeneracies at time reversal invariant momenta can form Kramers--Weyl (KW) fermions with quantized chiral charge \cite{Chang2018,Hasan2021}. Unlike conventional Weyl nodes, whose realization is generally material dependent and often associated with band inversion \cite{Hasan2021}, KW nodes are symmetry enforced at Kramers momenta and do not require band inversion. While cubic chiral materials such as the RhSi family have been widely explored for multifold chiral fermions \cite{Hasan2021,Sanchez2019,Rao2019}, scarce attention has been given to hexagonal chiral materials for exploration of multiple quantum geometry induced properties. SOC in these hexagonal chiral systems can stabilize twofold KW nodes with different quantized chiral charges \cite{Chang2018,Pu2026,Sato2025}. Notably, opposite structural enantiomers can preserve the same band energies while reversing the sign of the Berry curvature flux and the associated topological charges \cite{Pu2026}. Of particular importance is how the topological KW crossings are embedded within the Fermi surface, since a pocket enclosing a Weyl node acquires a nontrivial Berry flux and constitutes the low energy electronic structure accessible to quantum oscillations \cite{Alexandradinata2023,Emmanouilidou2020}.
	
Modern fermiology extends beyond the classical geometry of the Fermi surface to the quantum geometry of the Bloch states defined on it, thereby providing a route to probe the topological character of metallic electronic states \cite{Alexandradinata2023,Sato2025,Sharma2026,Kwan2020,Emmanouilidou2020}. In noncentrosymmetric chiral metals, antisymmetric spin orbit coupling (ASOC) \cite{Smidman2017supercondutivity} lifts the spin degeneracy away from the Kramers momenta, producing spin-split Fermi surfaces whose separation provides a direct measure of the underlying SOC energy scale \cite{Emmanouilidou2020,Sato2025}. Magnetic quantum oscillations are particularly sensitive to this low energy electronic structure. In the de Haas--van Alphen (dHvA) effect, frequency of each oscillation is determined by an extremal cross sectional area of the Fermi surface, while its angular and temperature dependences provide information on the pocket geometry  and cyclotron effective mass \cite{Alexandradinata2023,Sato2025}. Combined with first-principles calculations, the calculated dHvA branches can therefore be assigned to individual Fermi surfaces, allowing the spin-split pockets to be resolved and their relation to nearby KW crossings to be established. Emmanouilidou \textit{et al.}\cite{Emmanouilidou2020} validated the theoretical and experimental dHvA measurements within chiral NbGe$_2$  compound in $P6_222$ phase. Recently, Yao et al.\cite{Yao2025} experimentally observed chiral surface states in superconducting NbGe$_2$. Similarly, dHvA studies of the hexagonal VSi$_2$, NbSi$_2$, and TaSi$_2$ resolved the SOC split Fermi surfaces and quantified the characteristic energy scale of the ASOC \cite{Tsutsumi2013}.

Recent attention has increasingly focused on the role of the imaginary part of the quantum geometric tensor, encoded by the Berry curvature, in governing a wide range of electronic responses beyond conventional topological phases \cite{Gao2026QG,Liu2025QG}. In particular, the Berry curvature dipole (BCD) \cite{Sodemann2015,Ma2019,Joseph2024} gives rise to a second order nonlinear Hall response in time reversal symmetric noncentrosymmetric systems without need of magnetization or externally applied magnetic field, whereas the spin Berry curvature governs the intrinsic spin Hall conductivity (SHC) \cite{Sinova2015,Qiao2018,Dehghan2025}. The nonlinear Hall effect (NHE) has been widely studied in nonmagnetic materials including WTe$_2$ \cite{Du2018BCD_WTe2}, LiOsO$_3$ \cite{Xiao2020_BCDLiOsO3}, NbIrTe$_4$ \cite{Lee2024NbIrTe4}, TaAs \cite{Pang2024TaAs}, TaS$_2$ \cite{Wu2025TaS2}, Bi$_2$TeSe$_2$ \cite{Liao2026}, AsTeBr \cite{Rani2026}  and Ta$_2$Te$_2$O \cite{Sun2026}  although the results are often relatively weak or require precise symmetry breaking for finite values. Chiral crystals are therefore particularly attractive, as the absence of inversion and mirror symmetries relaxes constraints on the BCD tensor, allowing nonlinear Hall currents that can reverse sign between opposite enantiomers and also offer distinguishibility between it's pairs \cite{Joseph2024,Pu2026}. Furthermore, the spin Hall effect (SHE), quantified by SHC has potential applications in magnetic memories \cite{Amiri2024}, spin current injectors \cite{Montiel2023spin_injectors}, and spin transistors \cite{Guo2023_spin_transistors}. The intrinsic SHC is governed by the spin Berry curvature of the electronic bands and can be strongly enhanced near SOC induced avoided crossings and topological band degeneracies, where interband mixing becomes pronounced \cite{Sinova2015,Qiao2018,Dehghan2025}. Together, the BCD and SHC therefore provide complementary charge and spin transport signatures of the underlying quantum geometry.

Although hexagonal chiral structures provide a natural setting for topological electronic states, identifying highly metallic materials hosting numerous KW crossings with high Chern numbers and multiple Fermi surface sheets remains an important challenge. Previous studies of chiral C40 compounds, including TaGe$_2$, VSi$_2$, NbSi$_2$, NbGe$_2$ and TaSi$_2$ \cite{Gracia2021,Emmanouilidou2020}, have reported relatively few KW crossings owing to limited number of bands intersecting the Fermi level. Moreover, their associated topological surface states and Fermi arcs, together with the identification of Fermi surface pockets enclosing these topologically charged nodes, remains scarce. Furthermore, the possible enhancement of intrinsic spin Hall conductivity through energetic alignment with higher charge KW points has received little attention in this family. Similarly, the nonlinear Hall response remains largely unexplored within the chiral C40 family, with NbSi$_2$ being a only example \cite{Joseph2024}. These motivate further exploration of experimentally established C40 materials for large number of multicharged KW fermions, associated topological characteristics, quantum oscillations, multiple quantum geometric effects and the influence of structural handedness.
	
In the context of aforementioned fields, we investigate the experimentally synthesized chiral C40 aluminide WAl$_2$ using \textit{first-principles} calculations and symmetry analysis, considering both structural enantiomers. We systematically characterize its KW crossings through their topological charges, surface states, and Fermi arcs, and examine their relationship with the enclosing Fermi surface pockets. The underlying Fermiology is further investigated through angular dependent dHvA quantum oscillations, including signatures of ASOC Fermi surface splitting. We subsequently evaluate the Berry curvature dipole and associated nonlinear Hall transport. We have also examined the influence of crystal handedness on these quantum geometric responses and topological charges. Finally, we investigate the intrinsic spin Hall conductivity and its energetic alignment with higher charge KW crossings to understand their possible contributions to the spin Hall response.
	
	\begin{figure} 
		\centering
		\includegraphics[width=0.8\linewidth]{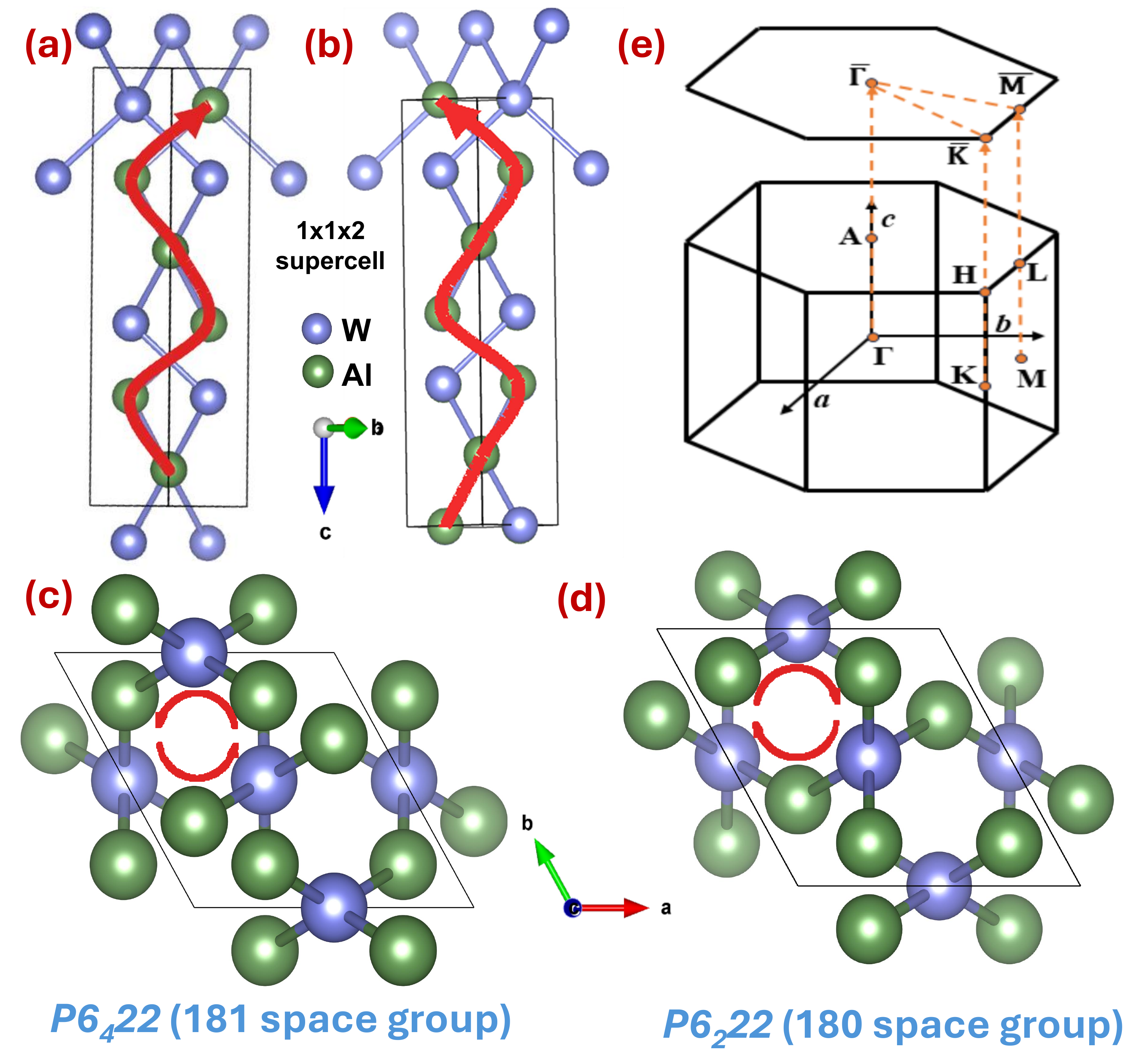}
		\caption{\justifying Side views of the optimized WAl$_2$ crystal structure in the two enantiomorphic chiral space groups, $P6_422$ (a) and $P6_222$ (b) respectively with a red arrows highlighting the opposite handedness of atomic arrangements along the \(c\) axis. To indicate the handedness, $1\times1\times2$ supercell is used in (a and b). The optimized structure perpendicular to the \(c\) axis for $P6_422$ (c)  and $P6_222$ (d) enantiomers, respectively. Here, W and Al atoms are shown as blue and green spheres, respectively. Bulk and surface hexagonal Brillouin zone along the (001) direction (e).}
		\label{Fig:structural+BZ}
	\end{figure}
	
	\begin{figure*}
		\centering
		\includegraphics[width=0.80\linewidth]{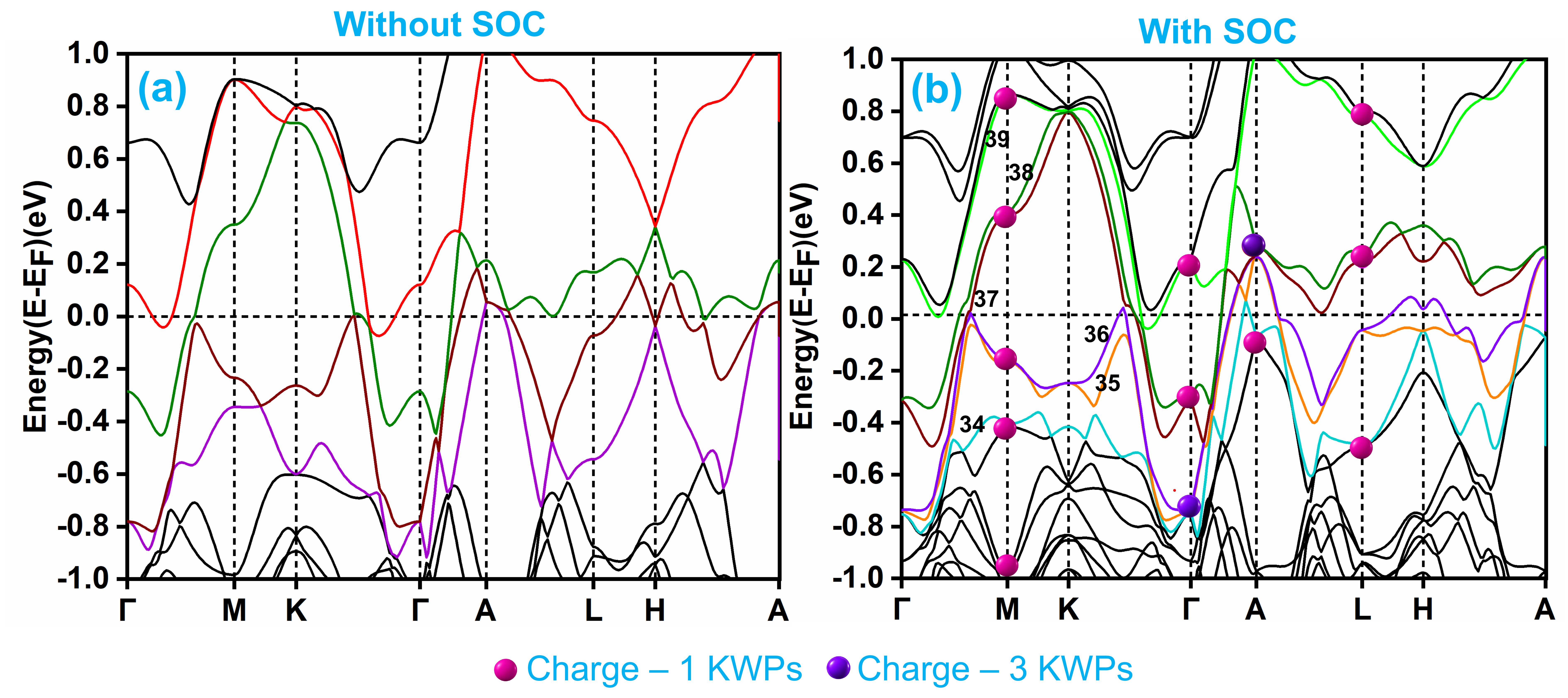}
		\caption{\justifying Non-relativistic (a) and relativistic (b) electronic band dispersion curves of WAl$_2$. The charge-1 and charge-3 Kramers--Weyl points are indicated by pink and purple spheres, respectively. Labels shown in panel (b) represents the band index near the Fermi level.}
		\label{fig:band}
	\end{figure*}

\section{\label{meth}Computational Methodology}
All calculations in the present work are performed using density functional theory (DFT) with projector augmented wave (PAW) method \cite{PAWBlochl1994} as implemented in the Vienna \textit{Ab initio} Simulation Package (VASP, version 6.4.2) \cite{VASPKresse1996}. The PAW formalism is used to describe the interaction between valence electrons and ionic cores, while the exchange correlation energy is treated using the Perdew-Burke-Ernzerhof (PBE) functional within the generalized gradient approximation (GGA) \cite{GGA}. A plane wave energy cutoff of $500\, \mathrm{eV}$ is used for both the phases, and Brillouin zone (BZ) is sampled using a $\Gamma$-centered $15\times15\times8$ $k$-point mesh during structural optimization for both the phases. The lattice parameters and atomic positions are fully relaxed during optimization until the residual Hellmann--Feynman forces on all atoms are below 0.001 eV \AA$^{-1}$ and the total energy convergence criterion of $10^{-8}\,\mathrm{eV}$ is satisfied. To analyze the topological nature of Kramers-Weyl fermions, electronic band structures of both the enantiomeric pairs are calculated with and without inclusion of relativistic effects. To investigate the thermal stability, ab initio molecular dynamics (AIMD) simulations are carried out at $T = 300\,\mathrm{K}$. Opimized geometries of the structures are visualized using VESTA \cite{Momma2011}. The tight binding Hamiltonian is generated within maximally localized Wannier functions (MLWFs) using \textsc{Wannier90} package~\cite{WANNIER90Mostofi2014} from a $d$-orbital of W atom whereas $s$- and $p$-orbitals of Al atom. Topological properties including chiral charge, surface states and Fermi arcs are analyzed with the \textsc{WannierTools} package~\cite{WANNIERTOOLS2018}.

The de Haas--van Alphen (dHvA) quantum oscillations are investigated using the Lifshitz--Kosevich (LK) theory \cite{Cambridge_Univ_Press_1984}, based on the Wannier interpolated band structure. For a given field direction, the magnetization oscillates as a sum over extremal orbit on the Fermi surface, which is given by \cite{Saini_PtAl_2022,Saini_YSi_2021};
\begin{equation}
	\Delta M \propto -B^k R_T R_D R_s
	\sin\!\left[2\pi\left(\frac{F}{B}+\gamma\right)\right].
\end{equation}
Here, $R_D$, $R_T$, and $R_S$ denote the Dingle, thermal, and spin damping factors, respectively, while $\gamma$ is the phase factor. The cyclotron effective masses are extracted from the thermal damping term, $R_T=X/\sinh X$, where $X=\lambda Tm^/B$ and $\lambda=2\pi^2 k_B m_e/(e\hbar)$. The effective mass $m^{*}$ of the carriers are expressed in units of the electron rest mass. Each orbit (say $i$) contributes a frequency $F_i$ related to the extremal cross sectional area of the Fermi surface by the Onsager relation  \cite{Onsager_1952,Alexandradinata2023};
\begin{equation}
	F_i = \frac{\hbar}{2\pi e}\,A.
\end{equation}
A $51\times51\times51$ $k$-point mesh was used for the Fermi surface analysis, with the extremal orbits $A$ and associated dHvA frequencies obtained using SKEAF \cite{SKEAF_2012}. All calculations are performed for magnetic fields along the $c$ axis, $H\parallel[001]$, and for field rotations toward the in-plane $\Gamma$--$M$ and $\Gamma$--$K$ directions over the angular range $\theta=-30^\circ$ to $60^\circ$ and the temperature range $1.8$--$3.2$~K.

The Berry curvature projections and the Berry curvature dipole (BCD) are computed with the  WannierBerri package \cite{WannierBerri_2021}. Within the Fermi sea formalism, the BCD is obtained using \cite{WannierBerri_2021};

\begin{equation}
	D_{ab} =
	\sum_n \int_{\mathrm{BZ}}
	\frac{d^3 k}{(2\pi)^3}
	f_n^0(\mathbf{k})
	\frac{\partial \Omega_n^b(\mathbf{k})}{\partial k_a}
	\, ,
	\label{eq:BCD_fermi_sea}
\end{equation}
where $f_n^0(\mathbf{k})$ is the equilibrium Fermi--Dirac distribution and $\Omega_n^b(\mathbf{k})$ is the Berry-curvature component of $n^{th}$ band. The indices $a$ and $b$ specify the cartesian directions in momentum space. The interpolated grid for the BCD calculation is 1440$\times$1440$\times$720. The corresponding nonlinear Hall response is evaluated within the Boltzmann transport framework using a finite relaxation time ($\tau$) \cite{Ziman2001,Rani2026}. The second order nonlinear Hall conductivity is related to the BCD through the Sodemann--Fu expression \cite{Sodemann2015,Joseph2024};
{\begin{equation}
		\chi_{abc} = -\epsilon_{adc}\,\frac{e^3\tau}{2\hbar^2(1+i\omega\tau)}\,D_{bd}.
		\label{nlhe}
\end{equation}
Here, \(\epsilon_{adc}\) represents the fully antisymmetric Levi-Civita symbol, \(e\) is the magnitude of the electron charge, \(\tau\) characterizes the relaxation time of charge carriers, and \(\omega\) denotes the angular frequency associated with the external electric field.
In the low frequency regime, where $\omega\tau \ll 1$, the system approaches the dc transport limit. Under this condition, Eq.~\ref{nlhe} simplifies to the expression, 
\begin{equation}
	\chi_{abc} = \varepsilon_{acd}\frac{e^3\tau}{2}D_{bd},
	\label{eq:nonlinear_Hall_conductivity}
\end{equation}
establishing a direct proportionality between the nonlinear Hall conductivity and the BCD tensor.

The intrinsic spin Hall conductivity (SHC) is evaluated within the linear response Kubo formalism \cite{Sinova_SHC_2004} using the spin-like Berry curvature. The SHC tensor component $\sigma^{\mathrm{s}}_{z,xy}$ is given by \cite{Qiao2018, Sinova_SHC_2004},

\begin{equation}
	\sigma^{\mathrm{s}}_{z,xy}
	=
	-e\sum_n\int_{\mathrm{BZ}}
	\frac{d^3k}{(2\pi)^3}
	f_{n\mathbf{k}}\,
	\Omega^{\mathrm{spin}}_{n,z,xy}(\mathbf{k}),
\end{equation}
where the spin Berry curvature associated with $n^{th}$ band index is defined as,
\begin{equation}
	\Omega^{\mathrm{spin}}_{n,z,xy}(\mathbf{k})
	=
	-2\,\mathrm{Im}
	\sum_{m\neq n}
	\frac{
		\langle n|j^{\mathrm{s}}_{z,x}|m\rangle
		\langle m|v_y|n\rangle
	}{
		(E_n-E_m)^2
	}.
\end{equation}
The SHC is calculated using $k$-mesh of $150\times150\times120$.

	\begin{figure*}
	\centering
	\includegraphics[width=0.8\linewidth]{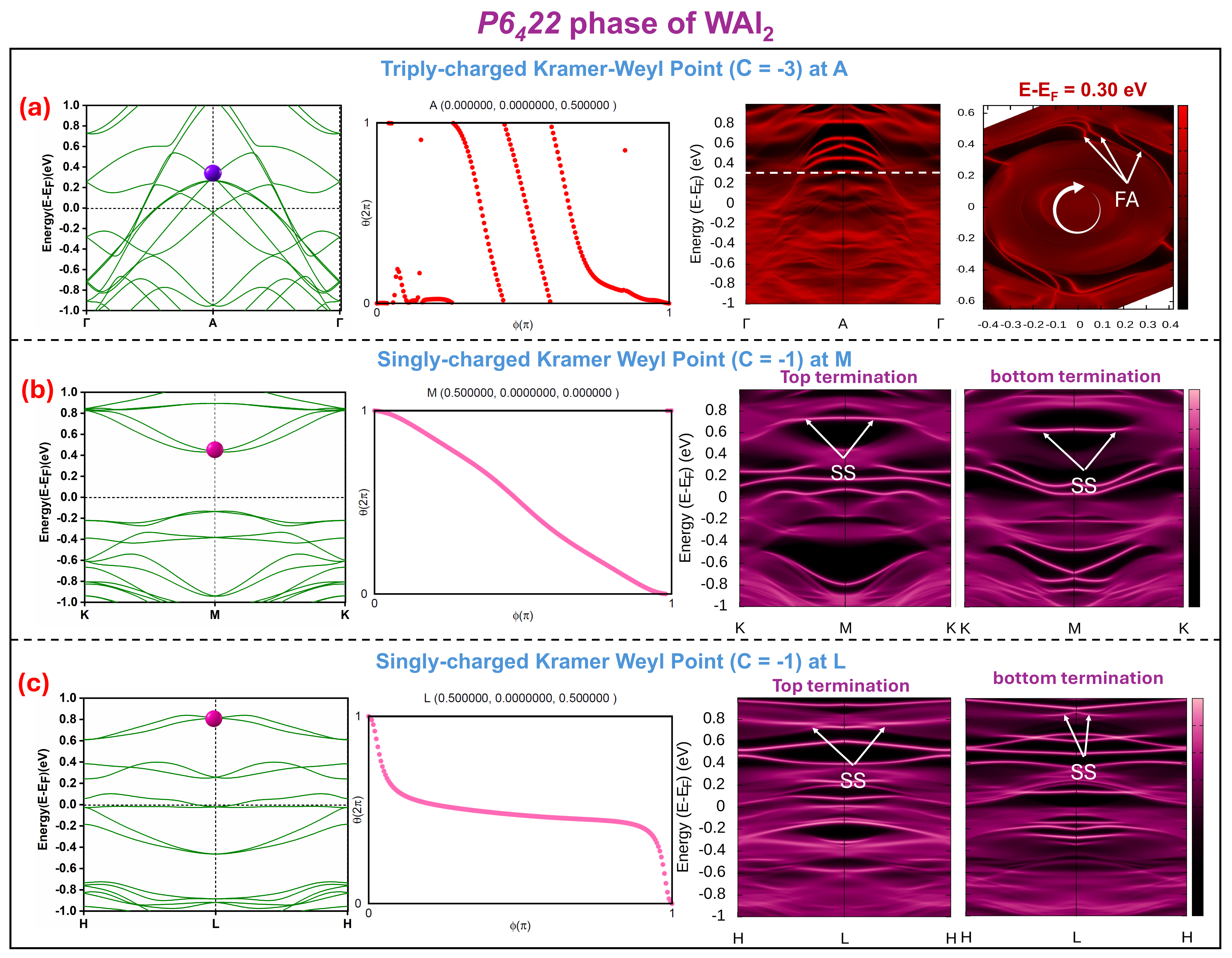}
	\caption{\justifying Triply charged Kramers--Weyl fermion (for $P6_{4}22$ phase) at the \(A\) point with chirality \(C=-3\) denoted by a purple sphere (a) in the bulk band structure along \(\Gamma\)--\(A\)--\(\Gamma\), the corresponding phase winding (WCC) plot, the surface spectral function showing the (010) surface in the energy range $-1$~eV to $+1$~eV with white arrows indicating the surface states and the Fermi arc at constant energy \(E-E_F=0.30\) eV. Singly charged Kramers--Weyl points at the \(M\) (b) and \(L\) (c) high-symmetry positions with chirality \(C=-1\) denoted by a pink sphere in bulk band structure along \(K\)-\(M\)-\(K\) (b) and  \(H\)-\(L\)-\(H\) (c) BZ directions. The corresponding phase winding (WCC) plot and surface spectral function for (010) surface, validates the \(C=-1\) chirality. The white circular arrow in panel (a) indicate the sense of circulation of the surface arc pattern.}
	\label{fig:topo_181}
\end{figure*}

\begin{figure*}
	\centering
	\includegraphics[width=0.80\linewidth]{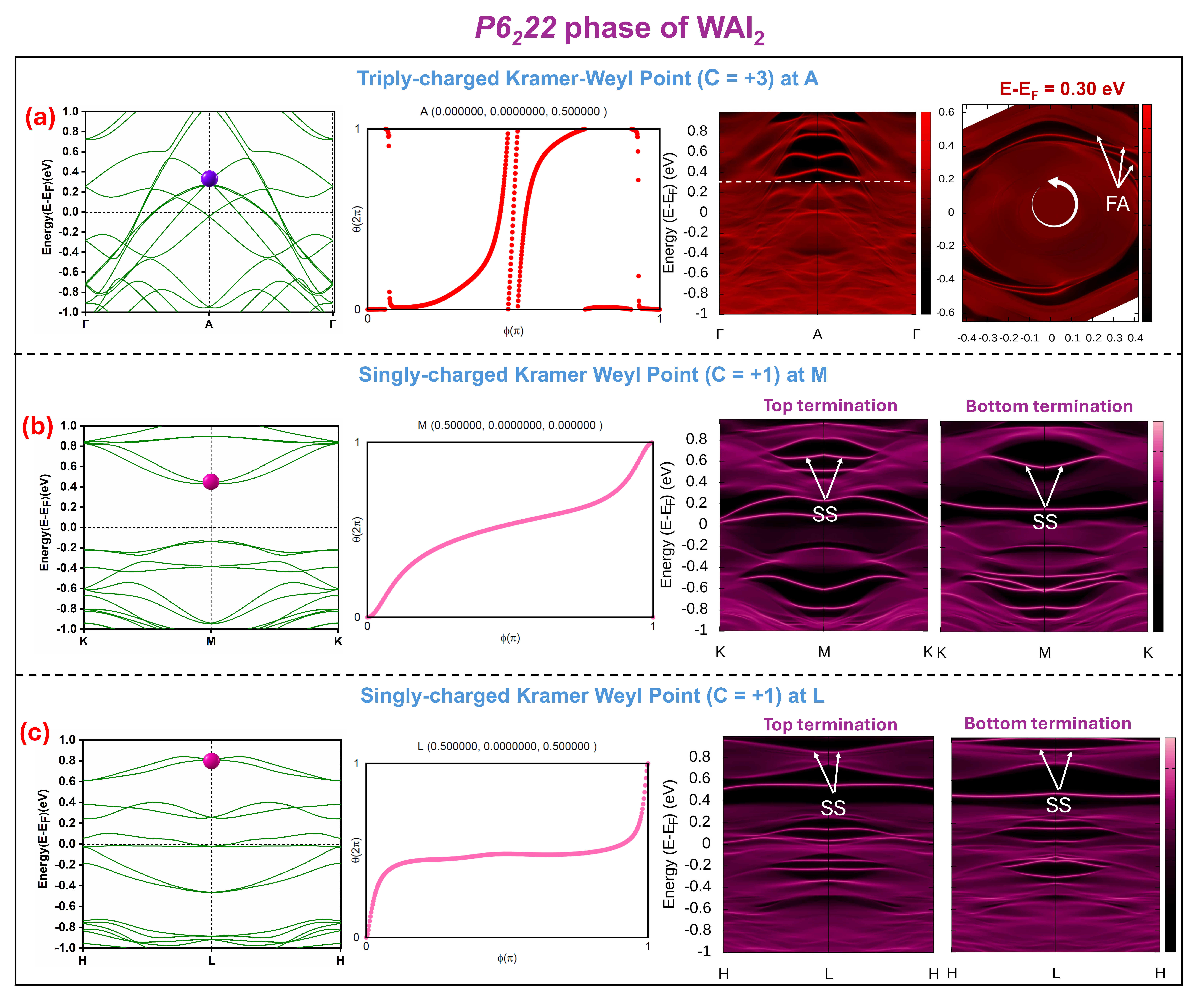}
	\caption{\justifying Triply charged Kramers--Weyl fermion (for $P6_{2}22$ phase) at the \(A\) point with opposite chirality \(C=+3\) denoted by a purple sphere (a) in the bulk band structure along \(\Gamma\)--\(A\)--\(\Gamma\). The corresponding phase winding (WCC) plot, surface spectral function for (010) surface with white arrows indicating surface states and the Fermi arc at constant energy \(E-E_F=0.30\) eV. The circular white arrow indicates the opposite sense of circulation than $P6_{4}22$ phase. Singly charged Kramers--Weyl points at the \(M\) (b) and \(L\) (c) high-symmetry positions with opposite chirality (\(C=+1\)) denoted by a pink sphere in bulk band structure along \(K\)-\(M\)-\(K\) (b) and  \(H\)-\(L\)-\(H\) (c) BZ directions. The corresponding phase winding (WCC) plot and surface spectral function for (010) surface, validates the \(C=+1\) chirality.}
	\label{fig:topo_180}
\end{figure*}
	
\section{Results and Discussions}
\subsection{Crystal Symmetry and Electronic structure}
WAl$_2$ is the first reported aluminide to crystallize in the $\mathrm{CrSi}_2$ type (C40) structure, as established by single crystal x-ray diffraction \cite{WAL2_2008_synthesis}. It crystallizes in the nonsymmorphic chiral space group $P6_{4}22$ (No.~181), characterized by hexagonal lattice and point group $622$ ($D_6$); the opposite structural enantiomorph belongs to $P6_{2}22$ (No.~180). Figure~\ref{Fig:structural+BZ}(a and b) illustrates the two enantiomorphic structures and the helical arrangement of W and Al atoms along the crystallographic $c$ axis, while the corresponding projections perpendicular to $c$ are shown in Fig.~\ref{Fig:structural+BZ}(c and d). The bulk and (001)-surface Brillouin zones (BZ) are shown in Fig.~\ref{Fig:structural+BZ}(e). In the optimized geometry, W occupies the symmetry-fixed $3d$ Wyckoff position $(\tfrac12,\tfrac12,\tfrac16)$, whereas Al occupies the $6j$ position $(x,1-x,\tfrac16)$, with the experimentally refined internal coordinate $x=0.8382(4)$ \cite{WAL2_2008_synthesis}. Accordingly, $x$ is the only independent internal structural parameter. Structural optimization yields the same lattice parameters for both enantiomorphs, with $a=4.86$~\AA{} and $c=6.79$~\AA{}, which are in good agreement with experimental parameters for $P6_{4}22$ phase ($a=4.74$~\AA{} and $c=6.60$~\AA{}). Notably, the calculated $c/a$ ratio remains essentially unchanged relative to experiment, with both values close to $1.39$.
	
The space group $P6_{4}22$ contains a $6_4$ screw axis parallel to $[001]$, combining a $60^\circ$ rotation with a translation of $\tfrac{2}{3}c$, together with twofold axes perpendicular to $c$-axis \cite{2005aCTA}. In the enantiomorphic $P6_{2}22$ phase, the corresponding $6_2$ screw axis combines the same rotation with a translation of $\tfrac{1}{3}c$, reversing the helical sense. Several C40 transition metal disilicides, including $\mathrm{VSi}_2$, $\mathrm{TaSi}_2$, and $\mathrm{NbSi}_2$, have been experimentally reported in the two enantiomorphic space groups discussed above \cite{Tanaka2001,2005aCTA}. More recently, chirality controlled growth has enabled the selective preparation of opposite handed $\mathrm{NbSi}_2$ and $\mathrm{TaSi}_2$ single crystals \cite{Kousaka2023}.
	
The chemical stability of WAl$_2$ is evaluated using its formation energy ($E_{\mathrm{form}}$)\cite{Saurav2025}. The calculated $E_{\mathrm{form}} = -0.07$~eV per formula unit indicates that the formation from elemental reference phases is energetically favorable, making decomposition into constituent elements energetically unfavorable. We further examine the thermal stability of WAl$_2$ using AIMD simulations. These simulations are carried out at $300$~K in the NVT ensemble using a $3\times3\times2$ supercell, with a time step of 2~fs over a total simulation time of 10~ps. As shown in Fig.~S1, the total energy exhibits only small fluctuations throughout the simulation, reflecting atomic vibrations around the equilibrium positions without noticeable structural distortion. These results support the thermal stability of WAl$_2$ in the $P6_{4}22$ phase at $300$~K.

The calculated electronic band dispersion with and without SOC for WAl$_2$ is elucidated in Fig.~\ref{fig:band}(a and b). The band structures of the two enantiomorphic phases exhibit identical dispersions, indicating that reversing the structural chirality preserves the overall band structure. The full BZ band dispersion for $P6_{2}22$ phase is not shown here. The band dispersions reveal the pronounced metallic character of WAl$_2$ relative to typical C40 transition metal disilicides \cite{Tanaka2001,2005aCTA,Kousaka2023,Emmanouilidou2020}, with bands number $34$--$39$ crossing the Fermi level along multiple high symmetry directions. Since the SOC is intrinsic part of all materials, hereafter we will discuss all properties with inclusion of SOC. All these bands are highlighted with different colors in Fig.~\ref{fig:band}. The SOC leads to the splitting of band degeneracies over the entire BZ except at TRIM ($\Gamma$, $M$, $A$, and $L$) points owing to the presence of all chiral symmetries. These symmetries also enforce only doubly degenerate band crossings at TRIM points. The states near the Fermi are dominated by the $5d$ orbitals of W atom, with a non negligible contribution from $s$ or $p$ orbitals of Al atom as shown in Fig. S2 of supplemental material \cite{supplemental_material}. The calculated density of states for WAl$_2$ is illustrated in Fig. S4 \cite{supplemental_material}. The pronounced metallicity reflected through high density of states (DOS) (\(N(E_F)\approx6.8\) states/eV) at the Fermi level have potential to facilitate the superconductivity \cite{Li2026_topo+super}. Recently, $\mathrm{ScSi}_2$ in $P6_{3}/mmc$ phase showcased enhanced electron--phonon coupling for superconductivity with \(N(E_F)\approx2.0\) states/eV \cite{Yang2026ScSi2}. We calculated DOS using \textsc{WannierTools} package~\cite{WANNIERTOOLS2018} with different broadening. Although these electronic features motivate the investigation of superconductivity in WAl$_2$, a quantitative evaluation of its electron--phonon coupling and superconducting transition temperature needs further investigation.

\begin{figure*}
	\centering
	\includegraphics[width=0.8\linewidth]{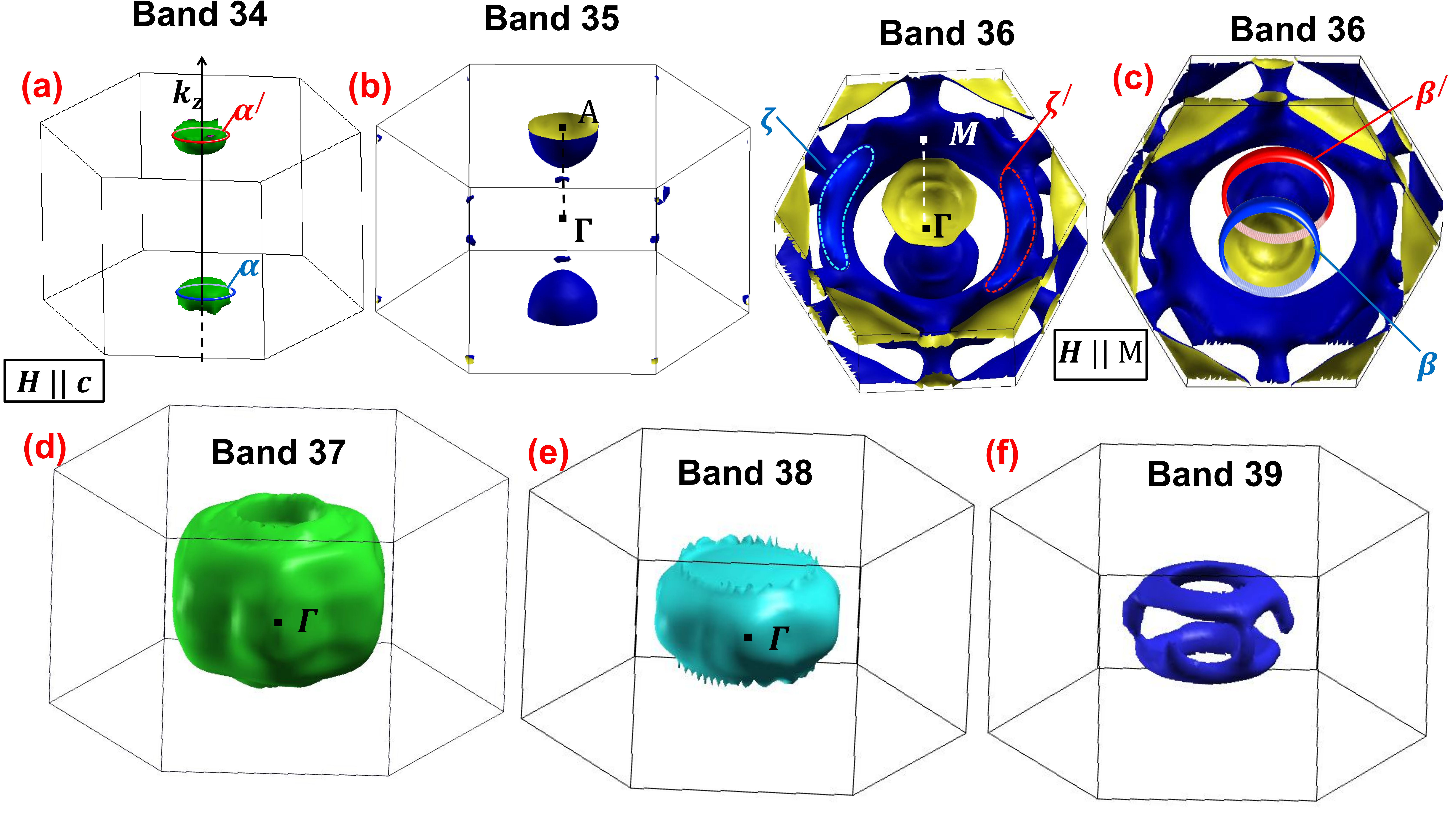}
	\caption{\justifying Fermi surface pockets of band 34, indicating two extremal frequencies forming an ASOC pair, $\alpha$ and $\alpha'$ along $k_z$ for magnetization $H \parallel c$ (a). Fermi surface pockets of band 35 enclosing the A point (b). Fermi surface for band 36 with ASOC pair of $\zeta$ and $\zeta'$ along $\Gamma \parallel M$ direction and ASOC pair of $\beta$ and $\beta'$ at the top and bottom of the Fermi surface for magnetization $H \parallel M$. Fermi surfaces of bands 37, 38 and 39, respectively with bands 37 and 38 forming closed pockets enclosing the $\Gamma$ point (d--f).}
	\label{fig:FS}
\end{figure*}

\subsection{Topological properties of Kramers-Weyl Fermions}
In nonmagnetic chiral crystals, the combination of time-reversal symmetry and SOC allows the Kramers degeneracies pinned to TRIM points to carry quantized Chern charge, giving rise to KW fermions \cite{Chang2018,Hasan2021}. In WAl$_2$, the relativistic band structure in Fig. \ref{fig:band}(b) reveals KW crossings at the $\Gamma$, $M$, $A$, and $L$ points \cite{Pu2026}. To further understand their topological character, we calculated the chirality of all degeneracies from -1 to 1 eV energy range using Wannier-charge-center evolution. The accuracy of the tight-binding is evaluated by comparing the interpolated band structure with \textit{first-principles} based DFT results (see Fig. S3 of the supplemental material\cite{supplemental_material}). The good agreement between both band structures validate the quality of Wannier90 generated tight-binding model\cite{Jay2025APL}.  The obtained topological charges at corresponding nodes are either $|C|=1$ or $|C|=3$ as represented with pink and purple color balls in Fig. \ref{fig:band}(b). The nodes at $M$ and $L$ points are singly charged KW fermions \cite{Fang2012}, whereas the crossings at $\Gamma$ and $A$ carry $|C|=3$ along with $|C|=1$ and therefore constitute higher charge KW fermions associated with the rotational and screw symmetries of the chiral lattice \cite{Fang2012,Tsirkin2017}. We observed twelve independent KW fermions with all six bands contributing to the Fermi surfaces.

In particular, we have analysed three representative KW degeneracies to further understand the surface states and Fermi arcs. Figure \ref{fig:topo_181}(a) shows the enlarged view of two-fold degeneracy at A point while the corresponding WCC evolution represents a threefold winding, identifying the node as a triply charged KW fermion for $P6_4 22$ phase. The associated surface spectral function shows dispersive surface states, and the constant energy surface at $E-E_F=0.30$~eV indicates three Fermi arcs emerging from $C=-3$.  In contrast, the representative crossings at the $M$ and $L$ points (see Fig.~\ref{fig:topo_181}(b and c)) exhibit single WCC winding and therefore carry unit topological charge. The calculated surface states for both terminations further validates the non-trivial topological nature.

The effect of structural handedness becomes evident upon comparing the two enantiomorphic phases. Although the electronic band dispersions are unchanged under reversal of the crystal chirality, the topological charges of the corresponding KW nodes reverse sign. As shown in Fig.~\ref{fig:topo_180}(a-c), the corresponding crossings remain at the same high symmetry momenta, but the WCC winding reverses direction, indicating reversal of the associated chiralities. The triply charged node at $A$ therefore changes from $C=-3$ in $P6_{4}22$ to $C=+3$ in $P6_{2}22$, while the singly charged nodes at $M$ and $L$ reverse from $C=-1$ to $C=+1$. The accompanying change in the circulation of the Fermi arc pattern (see Fig.~\ref{fig:topo_180}(a)) further reflects the reversal of momentum-space chirality between the two structural enantiomers. This type of topological chirality reversal is experimentally observed in cubic chirals such as PdGa \cite{2020_natcomm_PdGa}, CoSi and RhSn \cite{Li2019_CoSi}. The high symmetry locations, energies relative to the Fermi level, and topological charges of the KW nodes at the TRIM points for both enantiomorphs are summarized in Table~\ref{tab:chiral_table}. 
	
\begin{table}[htbp]
	\centering
	\caption{High-symmetry locations, energies relative to the Fermi level, and topological charges (chirality) of the Kramers--Weyl points at the TRIMs in both enantiomorphic phases of WAl$_2$.}
	\begin{ruledtabular}
		\begin{tabular}{ccccc}
			\textbf{KWP type} & \textbf{Position} & \textbf{$E-E_F$ (eV)} & \textbf{$P6_422$} & \textbf{$P6_222$} \\
			\midrule
			Triply charged KWP & $\Gamma$ & $-0.73$ & $+3$ & $-3$ \\
			Singly charged KWP & $\Gamma$ & $0.22$  & $+1$ & $-1$ \\
			Singly charged KWP & $\Gamma$ & $-0.31$ & $-1$ & $+1$ \\
			\midrule
			Triply charged KWP & $A$ & $0.27$  & $-3$ & $+3$ \\
			Singly charged KWP & $A$ & $-0.07$ & $-1$ & $+1$ \\
			\midrule
			Singly charged KWP & $M$ & $-0.41$ & $+1$ & $-1$ \\
			Singly charged KWP & $M$ & $-0.16$ & $+1$ & $-1$ \\
			Singly charged KWP & $M$ & $0.40$  & $-1$ & $+1$ \\
			Singly charged KWP & $M$ & $0.86$  & $+1$ & $-1$ \\
			\midrule
			Singly charged KWP & $L$ & $-0.48$ & $-1$ & $+1$ \\
			Singly charged KWP & $L$ & $0.23$  & $-1$ & $+1$ \\
			Singly charged KWP & $L$ & $0.79$  & $-1$ & $+1$ \\
		\end{tabular}
	\end{ruledtabular}
	\label{tab:chiral_table}
\end{table}

The different topological charges obtained at the TRIM points can be understood from the symmetries of the hexagonal chiral crystal. The chirality of each crossing is further constrained by the double valued irreducible representations (irrep) of its little group. At \(\Gamma\) and \(A\), the sixfold rotational symmetry allows distinct spinor representations with  both singly and triply charged KW nodes, whereas the spinor representation of the twofold symmetric little groups at $M$ and $L$ is associated with singly charged KW nodes ($|C|=1$) \cite{Fang2012,Chang2018,Tsirkin2017,Bradlyn2017,Pu2026}.  The detailed symmetry origin of the observed multicharged KWPs in WAl$_2$ is summarized in table \ref{tab:kw} and corresponding arguments is provided in section VI of supplemental material \cite{supplemental_material}. The  labels and representation follows Refs.~\cite{Bradley2009, Elcoro2017}, while the symmetry based chirality assignments follow Refs.~\cite{Fang2012,Chang2018,Tsirkin2017}. One can also fefer articles \cite{Pu2026} and \cite{Yu2022Encyclopedia} for more symmetry based arguments.

\begin{table*}[t]
	\caption{Symmetry based classification of the observed KW nodes at the TRIM points of WAl$_2$ including point groups, irreducible representations (irreps), symmetry generators and their matrix representations, generator eigenvalues, and topological chirality. Furthermore, detailed symmetry arguments are presented in section VI of supplemental material \cite{supplemental_material}}
	\label{tab:kw}
	\small
	\renewcommand{\arraystretch}{1.5}
	\begin{tabular*}{\textwidth}{@{\extracolsep{\fill}} c c c l c c @{}}
		\toprule
		\textbf{TRIM} & \textbf{Point group} & \textbf{Irrep.} &
		\textbf{Generator} &
		\textbf{Eigenvalues} & $\boldsymbol{|C|}$ \\
		\midrule
		$\Gamma$ & $D_6$ ($622$) & $R_7$ &
		$C_6^{+}\!:\ i\sigma_3$ &
		$\pm i$ & $3$ \\
		
		$\Gamma$ & $D_6$ ($622$) & $R_8$ &
		$C_6^{+}\!:\ (\sqrt3\sigma_0+i\sigma_3)/2$ &
		$e^{\pm i\pi/6}$ & $1$ \\
		
		$\Gamma$ & $D_6$ ($622$) & $R_9$ &
		$C_6^{+}\!:\ (-\sqrt3\sigma_0+i\sigma_3)/2$ &
		$e^{\pm i5\pi/6}$ & $1$ \\
		
		$A$ & $D_6$ ($622$) & $R_7$ &
		$C_6^{+}\!:\ i\sigma_3$ &
		$\pm i$ & $3$ \\
		
		$A$ & $D_6$ ($622$) & $R_8$ &
		$C_6^{+}\!:\ (\sqrt3\sigma_0+i\sigma_3)/2$ &
		$e^{\pm i\pi/6}$ & $1$ \\
		
		$A$ & $D_6$ ($622$) & $R_9$ &
		$C_6^{+}\!:\ (-\sqrt3\sigma_0+i\sigma_3)/2$ &
		$e^{\pm i5\pi/6}$ & $1$ \\
		
		$M$ & $D_2$ ($222$) & $R_5$ &
		$C_2\!:\ i\sigma_2$ &
		$\pm i$ & $1$ \\
		
		$L$ & $D_2$ ($222$) & $R_5$ &
		$C_2\!:\ i\sigma_2$ &
		$\pm i$ & $1$ \\
		\bottomrule
	\end{tabular*}
\end{table*}

\begin{figure*}
	\centering
	\includegraphics[width=0.80\linewidth]{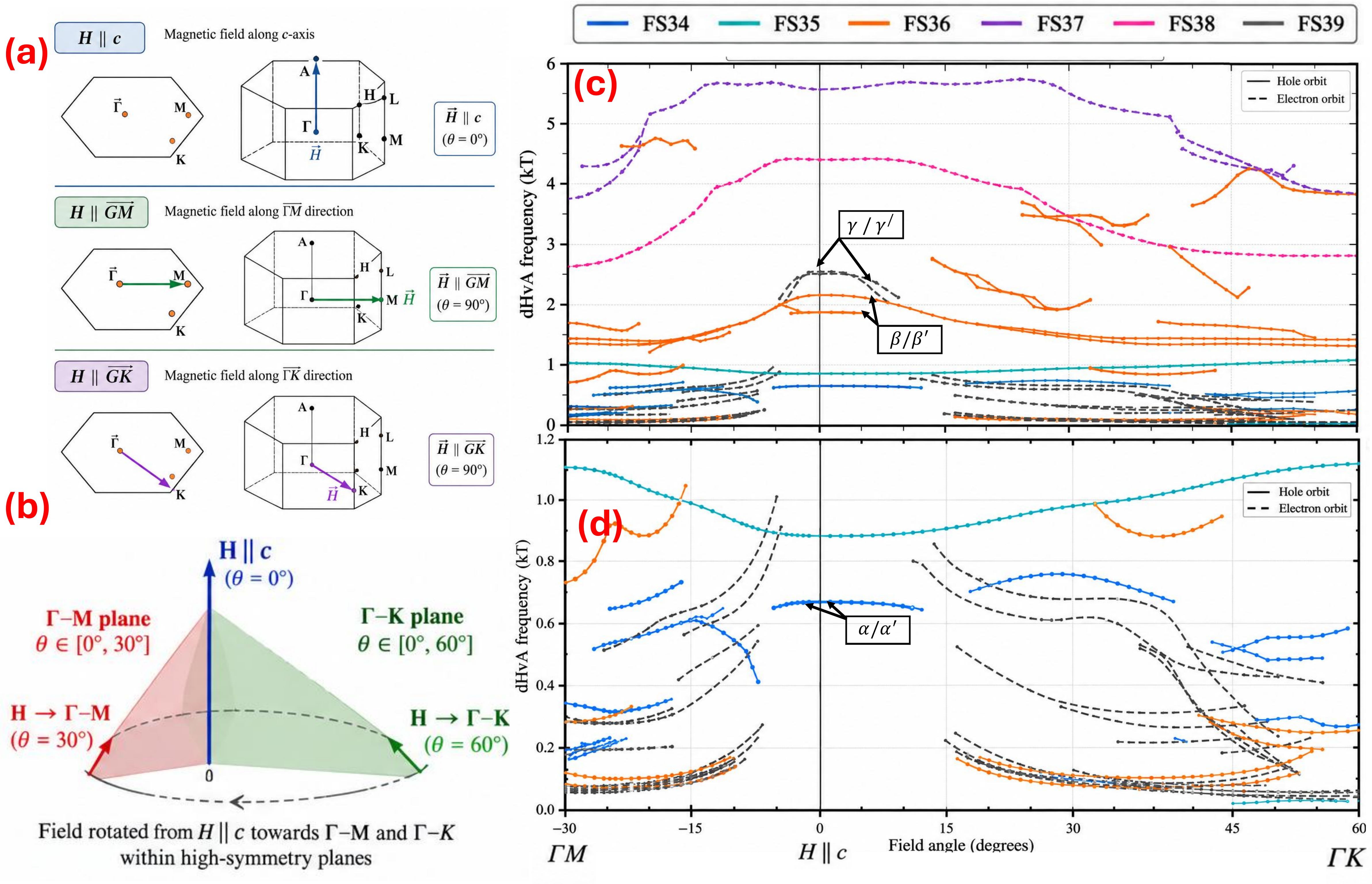}
	\caption{\justifying Schematic representation of the Fermi surface extremal orbit measurements for magnetic fields applied along the high symmetry directions \(\Gamma\!-\!M\), \(\Gamma\!-\!K\), and the \(c\) axis (a). Angular configuration of the magnetic field, illustrating its rotation from \(H\parallel c\) toward the \(\Gamma\!-\!M\) and \(\Gamma\!-\!K\) directions within the corresponding high symmetry planes (b). Angular dependence of the de Haas--van Alphen (dHvA) frequencies for the different Fermi surfaces (FS34--FS39) obtained by rotating the magnetic field within the \(\Gamma\!-\!M\) and \(\Gamma\!-\!K\) planes, respectively (c,d). Solid and dashed curves represent hole like and electron like extremal orbits, respectively. The labelled branches indicate the corresponding pairs of ASOC, including \(\gamma/\gamma'\), \(\beta/\beta'\), and \(\alpha/\alpha'\). }
	\label{fig:F_theta}
\end{figure*}	

\begin{figure*}
	\centering
	\includegraphics[width=0.80\linewidth]{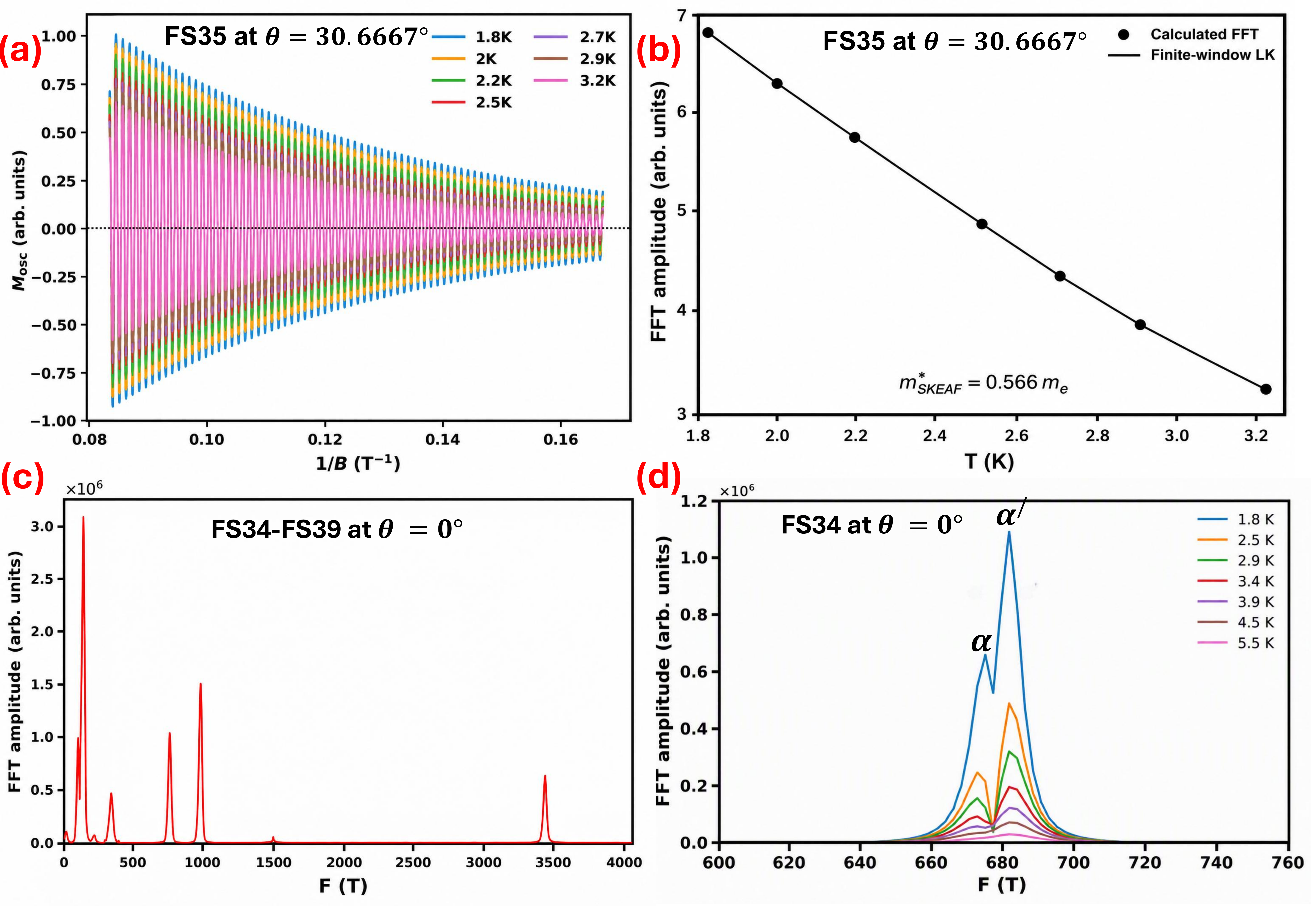}
	\caption{\justifying Oscillatory magnetization \(M_{\mathrm{osc}}\) vs \(1/B\) at different temperatures (a). The corresponding temperature dependence of the Fourier amplitude fitted using the finite window Lifshitz--Kosevich formalism, yielding $m_{\mathrm{SKEAF}}^{*}=0.566\,m_e$ (b). Fourier spectrum of \(M_{\mathrm{osc}}\) showing the observed quantum oscillation frequencies (c) and Fourier spectrum around the \(\alpha\) and \(\alpha'\) ASOC pairs at different temperatures for \(\theta\) = 0 (d).}
	\label{fig:LK}
\end{figure*}

\subsection{Fermiology and de Haas--van Alphen Effect}
The low-energy electronic structure of WAl$_2$ is characterized by six bands (band index 34-39 as illustrated in Fig.~\ref{fig:band}(b)) crossing the Fermi level giving rise to six distinct Fermi surface (FS) sheets.  As illustrated in Fig.~\ref{fig:FS}, the Fermi surface comprises three hole like sheets (FS34-FS36) and three electron like sheets (FS37-FS39). To investigate the detailed fermiology and quantum oscillation in WAl$_2$, we calculated the extremal cross sectional areas and corresponding de Haas--van Alphen (dHvA) frequencies using the Supercell K-space Extremal Area Finder (SKEAF) \cite{SKEAF_2012}. 

As evident from Fig.~\ref{fig:FS}(a-f), bands 34 and 35 form small hole pockets centered at the $A$ point, whereas band 36 gives rise to a larger, multiply connected hole sheet with several possible extremal cross sections. On the electron side, bands 37 and 38 form closed Fermi surface sheets around $\Gamma$, while band 39 produces smaller electron pockets in the vicinity of $\Gamma$.  Interestingly, most of these FSs are closed with distinct geometries which provides the basis for understanding the angular dependence and fermiology of the dHvA frequencies \cite{Muller_2020,Nguyen_2022,Wu_2025_RuO2}.

The angular dependence of the dHvA frequencies was examined by rotating the magnetic field within two high-symmetry planes, as illustrated in Fig.~\ref{fig:F_theta}(a). The field orientation is defined by the polar angle $\theta$, measured from the crystallographic $c$ axis, with $H\parallel c$ corresponding to $\theta=0^\circ$. Rotations toward the in-plane $\Gamma$--$M$ and $\Gamma$--$K$ directions were considered over the ranges $0^\circ\leq\theta\leq30^\circ$ and $0^\circ\leq\theta\leq60^\circ$, respectively, as shown in Fig.~\ref{fig:F_theta}(b). This angular dependent analysis describes the evolution of the extremal cross sectional areas and identify the anisotropic behaviour of the individual Fermi surface sheets \cite{Miyake2025,Brasse2013}. The similar approach is also applied for isostructural NbGe$_2$ for theoretical calculations by Emmanouilidou \textit{et. al.,} \cite{Emmanouilidou2020}. Their study also finds close agreement with experimental dHvA analysis.

The calculated angular dependence of the dHvA frequencies for rotations in the $\Gamma$--$M$ and $\Gamma$--$K$ planes is shown in Fig.~\ref{fig:F_theta}(c and d). The corresponding hole-like and electron-like orbits are represented by solid and dashed lines. Overall, the electron-like sheets FS37 and FS38 give rise to the prominent high-frequency branches, whereas the hole-like sheets FS34--FS36 contribute mainly to the lower-frequency region. The individual branches exhibit different angular dependences that reflect the geometry of their parent Fermi surface sheets. In particular, the branch associated with FS35 evolves smoothly with $\theta$, indicating a gradual change in its extremal cross sectional area as the magnetic field orientation is varied. In contrast, the complex FS36 sheet (Fig.~\ref{fig:FS}(c)) supports several extremal orbits whose cross sectional areas evolve differently with field direction, resulting in multiple frequency branches over the examined angular range. Several closely spaced frequency branches appear for selected magnetic field orientations, suggesting ASOC induced spin-splitting of the Fermi surfaces. We therefore further analysed corresponding extremal orbits to identify candidate ASOC split pairs and their frequency separations.

The candidate ASOC split pairs are identified by tracing closely spaced dHvA branches across the angular range. For $H\parallel c$, the $\alpha/\alpha'$, $\beta/\beta'$, and $\gamma/\gamma'$ pairs originate from FS34, FS36, and FS39, respectively, while additional pairs occur for field orientations H$\parallel$M and H$\parallel$K directions, mainly on FS36 (Fig.~\ref{fig:F_theta}). The corresponding low frequency angular evolution is illustrated in Fig. S4 of supplemental material \cite{supplemental_material}. The $\varepsilon/\varepsilon'$ pair is identified for $H\parallel M$ on the FS36 sheet (see Fig. S4 of supplemental material \cite{supplemental_material}). The corresponding ASOC splitting energy is estimated as $\Delta\varepsilon=(\hbar e/m_c^{*})|F_{-}-F_{+}|$, where $m_c^{*}$ is the cyclotron effective mass \cite{Miyake2025}. The candidate pairs, together with their frequencies, cyclotron effective masses, and estimated splitting energies, are summarised in Table~\ref{tab:ASOC_pairs}.  For $H\parallel c$, $\gamma/\gamma'$ shows the largest frequency separation, $26.3$~T, whereas $\alpha/\alpha'$ has the smallest, $4.0$~T. The cyclotron masses also vary between the different pairs, with $\delta/\delta'$ having the largest $m_c^{*}$. Since $\Delta\varepsilon$ depends on both the frequency separation and cyclotron mass, the resulting energy scale does not follow $\Delta F$ alone; for example, $\gamma/\gamma'$ gives $12.5$~K, while $\varepsilon/\varepsilon'$ yields a comparable value of $11.7$~K despite its smaller frequency separation. The calculated ASOC splitting energies in WAl$_2$ are of the same order as those reported for the isostructural chiral compound $\mathrm{VSi}_2$ \cite{Tsutsumi2013}.

\begin{table*}
	\centering
	\caption{\justifying Calculated dHvA frequency pairs associated with ASOC induced Fermi surface splitting in $WAl_2$ for selected magnetic field orientations. The table lists the frequencies ($F_{\pm}$), cyclotron effective masses ($m_{\pm}$), their average ($m_c^*$) and the corresponding splitting energy ($\Delta\varepsilon$).}
	\label{tab:ASOC_pairs}
	\begin{ruledtabular}
		\begin{tabular}{ccccccccc}
			\textbf{Pair} & \textbf{Field} & \textbf{Sheet} & \textbf{$F_{-}$ (T)} & \textbf{$F_{+}$ (T)} &
			\textbf{$m_{-}$ ($m_e$)} & \textbf{$m_{+}$ ($m_e$)} & \textbf{$m^{*}_{c}$ ($m_e$)} &
			\textbf{$\Delta\varepsilon$ (K)} \\ [4pt]
			\hline \\ [4pt]
			$\alpha/\alpha'$ & $H\!\parallel\! c$ & FS34 & 676.5 & 680.5 & 0.830 & 0.757 & 0.794 & 6.8 \\[4pt]
			$\beta/\beta'$   & $H\!\parallel\! c$ & FS36 & 1889.5 & 1897.4 & 2.5382 & 2.6786 & 2.608 & 4.1 \\[4pt]
			$\gamma/\gamma'$ & $H\!\parallel\! c$ & FS39 & 2522.8 & 2549.1 & 2.5927 & 3.0784 & 2.836 & 12.5 \\[4pt]
			$\delta/\delta'$ & $H\!\parallel\! K$ & FS36 & 3855.6 & 3885.0 & 5.5666 & 5.5315 & 5.549 & 7.1 \\[4pt]
			$\varepsilon/\varepsilon'$ & $H\!\parallel\! M$ & FS36 & 724.9 & 736.3 & 1.3047 & 1.3051 & 1.305 & 11.7 \\[4pt]
			$\zeta/\zeta'$   & $H\!\parallel\! M$ & FS36 & 1057.9 & 1066.9 & 1.6855 & 1.7135 & 1.700 & 7.1
		\end{tabular}
		\end{ruledtabular}
	\end{table*}
	
Beyond identifying the extremal orbit frequencies and cyclotron effective masses, the present analysis uses the SKEAF-derived quantities to construct a theoretical Lifshitz--Kosevich (LK) \cite{Cambridge_Univ_Press_1984} description of the quantum oscillation response of WAl$_2$. The temperature dependence of the oscillation amplitude provides a means of examining the cyclotron effective mass associated with an extremal orbit. Figure~\ref{fig:LK}(a) shows the calculated oscillatory magnetization for a representative hole-like orbit of FS35 over the field range $2$--$35$~T and temperatures between $1.8$ and $3.5$~K. The oscillations are periodic in $1/B$, as expected from the Landau quantization of extremal Fermi surface orbits \cite{Cambridge_Univ_Press_1984,Emmanouilidou2020,Alexandradinata2023}. The temperature dependence of the oscillation amplitude is governed by the LK thermal damping factor $R_T$ introduced in Sec.~II. For the FS35 orbit at $\theta=30.67^\circ$ in the $\Gamma$-K rotation plane, the temperature dependence is evaluated over $1.8$--$3.2$~K using the  cyclotron mass $m_c^*=0.57\,m_e$, as shown in Fig.~\ref{fig:LK}(b). The Fourier transform for $H\parallel c$, shown in Fig.~\ref{fig:LK}(c), displays the frequency peaks associated with the calculated extremal orbits. These peaks resolve the individual oscillation frequencies and connect the simulated response to the corresponding extremal Fermi surface areas through the Onsager relation \cite{Cambridge_Univ_Press_1984,Emmanouilidou2020,Alexandradinata2023}. Figure~\ref{fig:LK}(d) shows the temperature evolution of the Fourier \textcolor{blue}{amplitudes} associated with the candidate $\alpha$ and $\alpha'$ branches from $1.8$ to $5.5$~K. The two closely spaced components remain distinguishable in the Fourier spectrum while their amplitudes decrease with increasing temperature, providing a direct representation of the expected thermal evolution of the candidate ASOC split branches. These calculations provide theoretical predictions for the quantum oscillation signatures of WAl$_2$ and a basis for comparison with future experimental measurements.

\begin{figure*}
	\centering
	\includegraphics[width=0.85\linewidth]{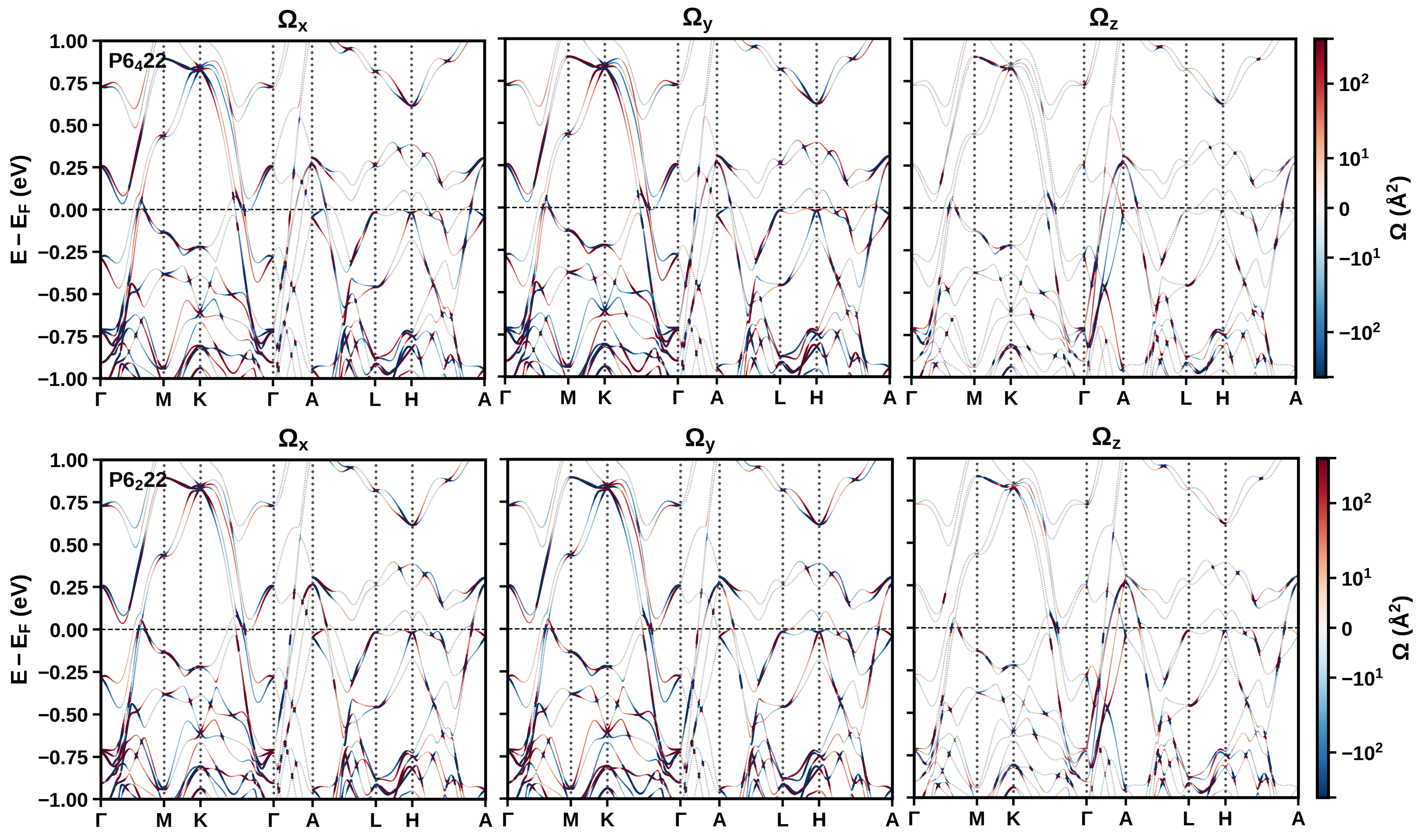}
	\caption{\justifying Momentum resolved berry curvature in $P6_422$ (a) and $P6_222$ (b) respectively, projected on the full BZ. Here, three columns display the $x$, $y$ and $z$ components of the Berry curvature vector, respectively. The two enantiomers display opposite concentration of Berry curvature in all three components. Strongly localized Berry curvature features are particularly prominent around band crossing regions.}
	\label{fig:berry_band}
\end{figure*}

\subsection{Origin of Intrinsic Berry Curvature Dipole}
	
In general, BCD originates from the asymmetric Berry curvature distribution owing to lack of inversion symmetry in materials \cite{Xiao2010}. BCD is defined as the first moment of the Berry curvature distribution over the occupied electronic states in momentum space as described in equation \ref{eq:BCD_fermi_sea}. The BCD may also be viewed as a nonlinear analogue of the Drude weight, governing the nonlinear Hall response in much the same way that the conventional Drude weight characterizes linear electrical conductivity \cite{Matsyshyn2019,Zeng2021}. The twofold rotational symmetries of the $622$ ($D_6$) point group force the off-diagonal BCD components to vanish, while the sixfold rotational symmetry relates the in-plane diagonal components, giving $D_{xx}=D_{yy}$ and $D_{zz}=-2D_{xx}$ \cite{Joseph2024}. The momentum resolved Berry curvature, overlaid on the electronic band structure along the $\Gamma$--M--K--$\Gamma$--A--L--H--A path, is presented in Fig.~\ref{fig:berry_band}(a and b) for $P6_422$ and $P6_222$ phases. The Berry curvature exhibits pronounced enhancements near band crossings and anticrossings, reaching values on the order of $10^{2}$~\AA $^{2}$ after normalization. This strong asymmetric Berry curvature distribution relative to previously reported materials suggests large nonlinear Hall effect (NHE) \cite{Sun2026,Zhou2026,Bandyopadhyay2022,Joseph2024}.  Among the three components, the out of plane component $\Omega_z$ is comparatively less pronounced. Notably, all three Berry curvature components reverse sign between the two enantiomers, reflecting their opposite structural handedness.

The BCD tensor can be written as $D=D_{xx}\,\mathrm{diag}(1,1,-2)$, with its overall sign determined by the structural handedness \cite{Sodemann2015,Joseph2024}. Figure~\ref{fig:bcd}(a and b) shows the corresponding nonzero BCD components, $D_{xx}$, $D_{yy}$, and $D_{zz}$, as functions of the Fermi energy for the two enantiomers. The calculated components satisfy the symmetry relations $D_{xx}=D_{yy}$ and $D_{zz}=-(D_{xx}+D_{yy})=-2D_{xx}$ throughout the energy range considered in the vicinity of Fermi energy. At the Fermi level, the in-plane component has a magnitude of $|D_{xx}|\approx0.027$, while the out-of-plane component reaches $|D_{zz}|\approx0.054$. Consistent with the reversal of the Berry curvature distribution, the BCD components change sign between the two enantiomorphic phases (see Fig.~\ref{fig:bcd}(a and b)).

The magnitude and energy dependence of the BCD are governed by the momentum space distribution of the Berry curvature \cite{Zeng2021}. In WAl$_2$, large Berry curvature contributions arise near SOC induced avoided crossings and topological degeneracies of the band structure, which consequently produce pronounced features in the BCD. The BCD reversal is consistent with the reversal of the topological chiralities discussed in Sec.~[B].

\begin{figure*}
	\centering
	\includegraphics[width=0.80\linewidth]{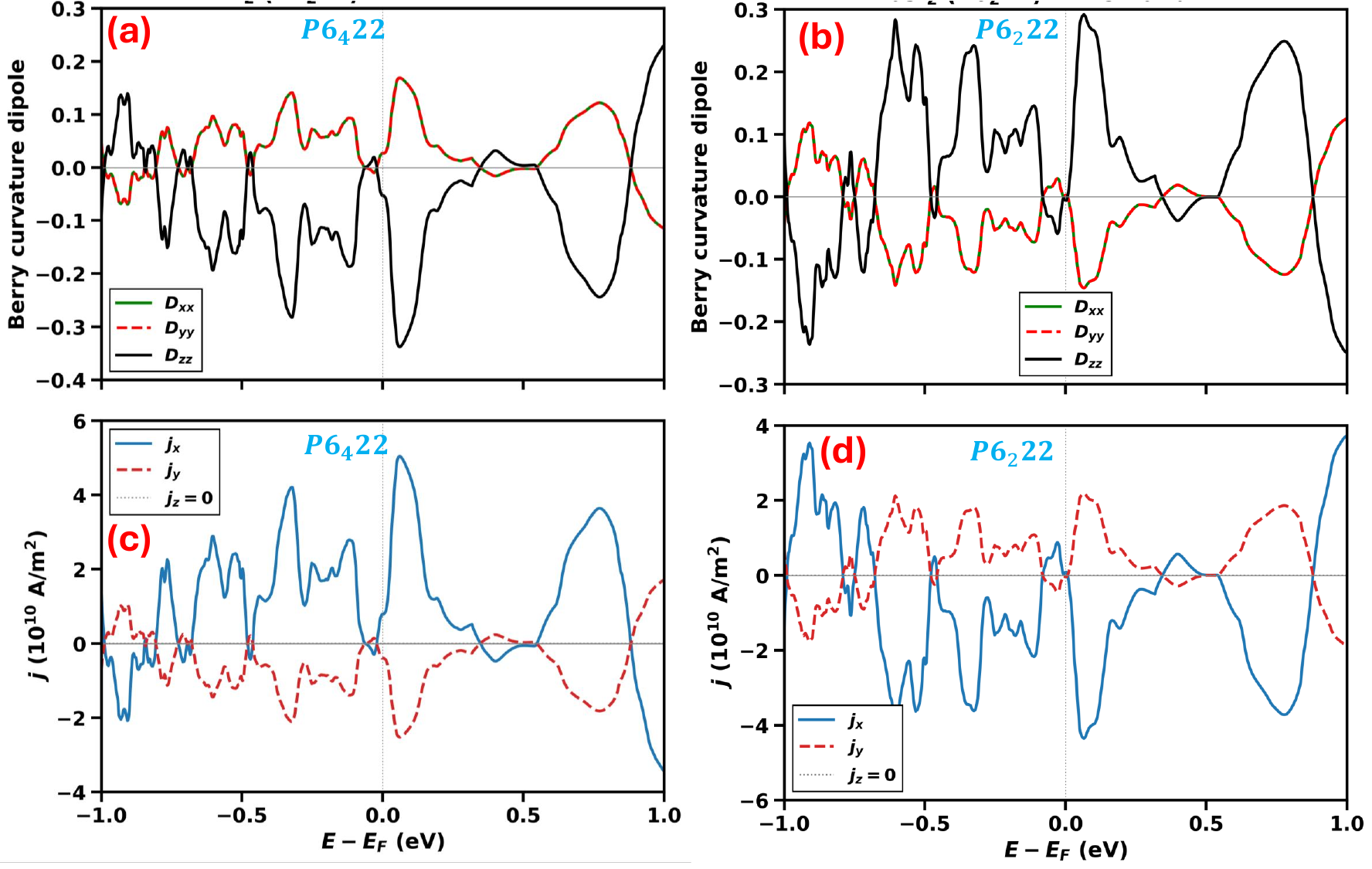}
	\caption{\justifying
		The upper panel shows the diagonal components of the Berry curvature dipole, $D_{xx}$, $D_{yy}$ and $D_{zz}$, while the lower panel represents the corresponding second order nonlinear current densities, $j_x$ and $j_y$, with $j_z=0$. Panels (a and c) correspond to the $P6_4 22$ phase, whereas panels (b and d) correspond to $P6_2 22$ phase. The Fermi level is set to zero.
	}
	\label{fig:bcd}
\end{figure*}

\subsection{Nonlinear Hall Response}

The finite Berry curvature dipole of WAl$_2$ generates a nonlinear Hall response \cite{Sodemann2015}. Under time-reversal symmetry, the Berry-curvature contributions from opposite momenta cancel at equilibrium. When an electric field is applied, the Fermi surface distribution becomes tilted, producing a finite net Berry curvature and thereby enabling a nonlinear Hall response \cite{Gao2026QG}. The large finite BCD observed in WAl$_2$ leads to the nonlinear Hall response according to equation \ref{eq:nonlinear_Hall_conductivity}. The resulting transverse current is quadratic in the field and is given by; \cite{Arka2024,Sodemann2015}:
\begin{equation}
	j_\alpha = \chi_{\alpha\beta\gamma} E_\beta E_\gamma,
	\label{eq:nlhe}
\end{equation}
where $\chi_{\alpha\beta\gamma}$ is the second order conductivity tensor. We evaluate the in-plane current densities $j_x$ and $j_y$ generated by an applied electric field, while out-of-plane component $j_z$ vanishes due to symmetry constraints \cite{Joseph2024}.

In a typical transport geometry, the relaxation time is of the order of picoseconds, whereas the driving frequency lies in the $10$--$1000~\mathrm{Hz}$ range \cite{Ma2019,Kang2019}. Consequently, the frequency dependent factor in the nonlinear conductivity tensor $\chi$ can be neglected \cite{Roy2022,Saha2023,Joseph2024}. Here, $j_i$ is driven by the electric field components perpendicular to the current direction, each evaluated as $E_i=V_i/L_i$. For a representative crystal of dimensions $10\times5\times3~\mu\mathrm{m}^3$, assuming an isotropic resistivity of $\rho=40~\mu\Omega\,\mathrm{cm}$ and an applied voltage of $V=2~\mathrm{V}$ across each pair of opposite faces, the estimated current densities are $j_x=2.98\times10^{11}D_{xx}~\mathrm{A\,m^{-2}}$ and $j_y=-1.49\times10^{11}D_{xx}~\mathrm{A\,m^{-2}}$, corresponding to voltage drops of approximately $1.2D_{xx}~\mathrm{V}$ and $0.30D_{xx}~\mathrm{V}$. A relaxation time of \(\tau = 3\) ps was used in the estimation of the nonlinear Hall current densities. These estimates are comparable to those reported for the isostructural compound $\mathrm{NbSi}_2$ by Joseph \textit{et al.}~\cite{Joseph2024} and are within the current experimental capabilities. Furthermore, the magnitude of the BCD component is $|D_{xx}|=0.027$ at the fermi level which corresponds to in-plane current densities of $|j_x|\approx 7.95\times10^{9}\,\mathrm{A\,m^{-2}}$ and $|j_y|\approx 3.98\times10^{9}\,\mathrm{A\,m^{-2}}$. The energy dependence of the second-order Hall response is shown in Figs.~\ref{fig:bcd}(c and d) for both phases of WAl$_2$. Both $j_x$ and $j_y$ closely follow the variation of the BCD, exhibiting pronounced peaks and sign changes as the Fermi level passes through regions of strong Berry curvature contribution. Most importantly, the current response reverses between the two enantiomers, directly reflecting the chirality dependent reversal of the BCD, while $j_z$ remains zero by symmetry. Thus, the nonlinear Hall current provides a direct transport signature of the structural handedness in WAl$_2$. Notably, the enhanced BCD gives rise to a stronger nonlinear Hall response. At $E-E_F=+0.06$~eV, the BCD reaches $D_{xx}=D_{yy}\approx0.169$ and $D_{zz}\approx-0.338$. Consistent with this enhancement, the nonlinear Hall current also attains its maximum magnitude at the same energy, with $|j_x|\approx5.04\times10^{10}~\mathrm{A\,m^{-2}}$ and $|j_y|\approx2.52\times10^{10}~\mathrm{A\,m^{-2}}$. Hence, slight Fermi level tuning in WAl$_2$ significantly enhance the Nonlinear response. 

As dipole vanishes after the restoration of the relevant symmetries, it can also probe ferroelectric and topological phase transitions \cite{Xiao2020,Facio2018,Sinha2022}. The nonlinear Hall effect has now been experimentally reported in several classes of quantum materials, including van der Waals systems \cite{Ma2019,Kang2019}, topological crystalline insulators \cite{Zhang2022_ferro_bcd, Nishijima2023}, and Dirac and Weyl semimetals \cite{Kumar2021,Shvetsov2019}. In addition to its fundamental relevance, the effect has attracted interest for possible use in terahertz detection \cite{Zhang2021_THz_NHE} and nonlinear electronic devices \cite{Ramsden2011}.

\subsection{Spin Hall Conductivity and role to Kramers-Weyl Fermions}
Generating spin currents efficiently is an important requirement for spin-based electronic technologies \cite{Dehghan2025,Zhang2021SHC}. One of the key mechanisms for achieving this is the intrinsic spin Hall effect, which is governed by the SOC coupled electronic structure of the material rather than by impurity scattering \cite{Sinova2015, SHC_topo_2023}. The resulting transverse spin response is closely connected to the spin Berry curvature of the electronic bands, which gives rise to an anomalous carrier velocity under an applied electric field \cite{Sinova2015}. Recent work by Ascencio \textit{et al.}\cite{SHC_topo_2023} has shown that the Kramers--Weyl fermions can substantially enhance the intrinsic spin Hall conductivity in cubic chiral topological semimetals. Motivated by this connection, it is pertinent to ask whether high chern number KW fermions enhance the SHC in hexagonal chiral also?  we therefore investigate the intrinsic spin Hall response of WAl$_2$.

For the $622$ point group of WAl$_2$, crystal symmetry restricts the axial rank-three spin Hall conductivity tensor to six nonzero components, of which only three are independent \cite{Dehghan2025}. Here, we focus on the conventional component $\sigma_{xy}^{z}$, which describes a spin current flowing along the $x$ direction with spin polarization along $z$, generated by an electric field applied along $y$.

To examine the microscopic origin of this response, the momentum resolved spin Berry curvature is projected onto the electronic bands along the same high symmetry path used for the band structure, as shown in Fig.~\ref{fig:SHC}(a). Pronounced spin Berry curvature contributions appear near several band crossings and SOC modified anticrossings. Such regions are expected to contribute strongly to the intrinsic spin Hall conductivity because the latter is obtained by integrating the spin Berry curvature of the occupied states over the BZ \cite{Sinova2015,SHC_topo_2023}. The corresponding energy dependent $\sigma_{xy}^{z}$, shown in Fig.~\ref{fig:SHC}(b), exhibits substantial variation with the chemical potential. At the Fermi level, $\sigma_{xy}^{z}$ is approximately $-160.30\,(\hbar/e)\,\mathrm{S\,cm^{-1}}$. Within $\pm0.5$~eV of the Fermi level, its magnitude increases to $236.24\,(\hbar/e)\,\mathrm{S\,cm^{-1}}$ at $E-E_F=-0.48$~eV, indicating that the spin Hall response is sensitive to the electronic states sampled by the chemical potential.

A still larger response, $|\sigma_{xy}^{z}|\approx254.83\,(\hbar/e)\,\mathrm{S\,cm^{-1}}$, occurs at $E-E_F=-0.70$~eV. Interestingly, this energy lies close to the triply charged Kramers--Weyl point at $\Gamma$, located at $E-E_F=-0.72$~eV as listed in Table~\ref{tab:chiral_table}. The enhancement of spin Berry curvature in this energy region suggests that states associated with the higher charge Kramers--Weyl crossing contribute to the increased spin Hall response. Nevertheless, because the SHC represents an integral over the entire Brillouin zone, the enhanced value in WAl$_2$ should be regarded as the combined contribution of the Kramers--Weyl states and other nearby spin Berry curvature hot spots rather than being assigned exclusively to the $\Gamma$ point.

For comparison, the calculated Fermi level value in WAl$_2$ is significantly larger in magnitude than those reported for several related chiral semimetals, including CoSi ($-67$), CoGe ($-85$), RhSi ($-113$), RhGe ($-139$) \cite{Hsieh2022}, TaSi$_2$ ($-146$), NbSi$_2$ ($-173$) and VSi$_2$ ($114$) \cite{Zhang2021SHC,Dehghan2025}. All values are in unit of $(\hbar/e)\,\mathrm{S\,cm^{-1}}$. Together with the pronounced energy dependence of $\sigma_{xy}^{z}$, this comparison indicates that the spin Hall response of WAl$_2$ is appreciable among chiral semimetals and can be further enhanced when the chemical potential approaches regions of large spin Berry curvature.

\begin{figure*}
	\centering
	\includegraphics[width=0.8\linewidth]{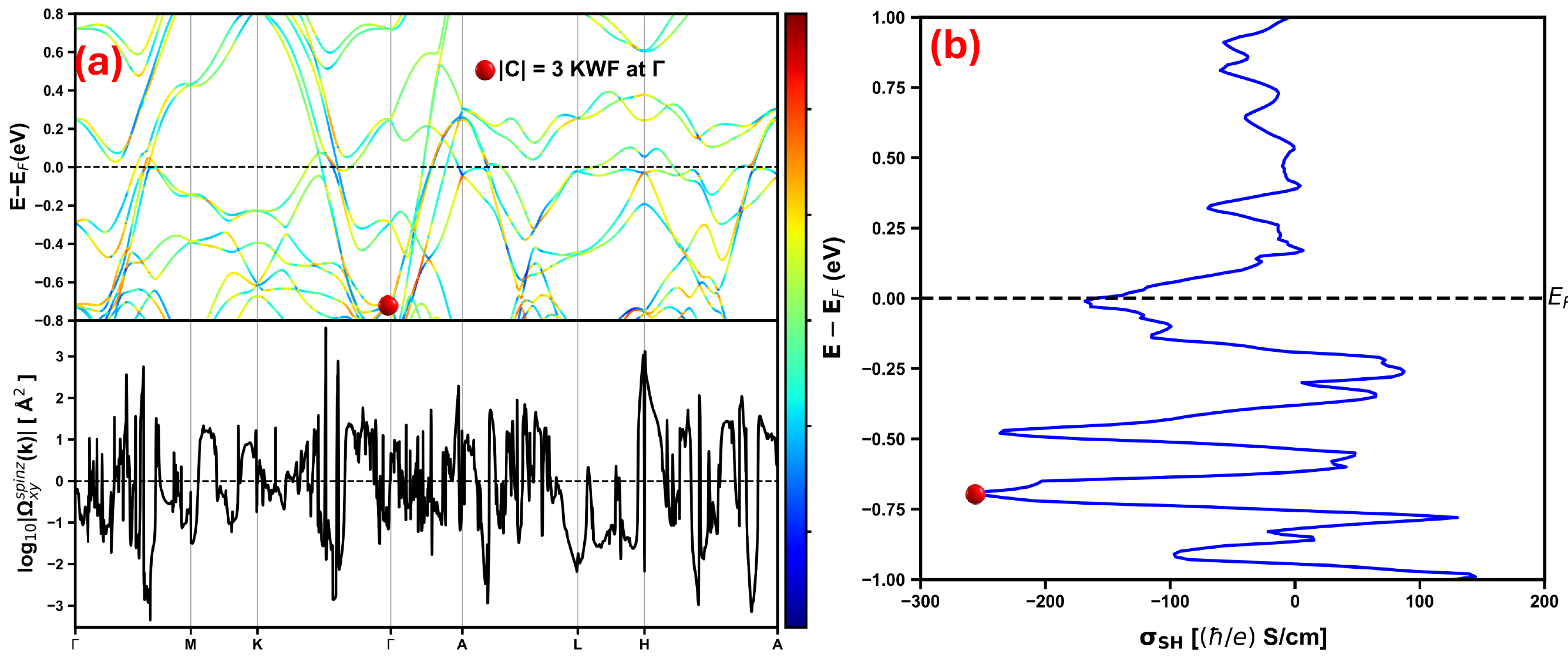}
	\caption{\justifying Momentum resolved spin Berry curvature projected (a) onto the electronic structure along the full BZ (top) together with corresponding logarithmic scale representation of the spin Berry curvature (bottom). The calculated spin Hall conductivity of WAl$_2$ (b). The red sphere indicates the maximum value of SHC aligned with the triply charged $|C|=3$ KWF at $\Gamma$.}
	\label{fig:SHC}
\end{figure*}
	
\section{Conclusions}
In summary, we investigated Kramers--Weyl (KW) topology, multiband fermiology, de Haas--van Alphen (dHvA) quantum oscillations and quantum geometric Hall responses in WAl$_2$ using \textit{first-principles} calculations and symmetry analysis. We identify \textit{twelve} KW crossings with chiral charges $|C|=1$ and $|C|=3$, which is significantly higher than previously reported chiral C40 compounds. Their nontrivial topological character is further supported by WCC evolution, surface states, and Fermi arcs. Symmetry analysis further resolves distinct KW chiral charges through their underlying little group representations. The six electron and hole like Fermi surfaces exhibit characteristic extremal orbits, with six candidate ASOC frequency pairs resolved in the calculated dHvA calculations. Their angular dependence and Lifshitz--Kosevich behavior provide specific predictions for experimental Fermiology.

The quantum geometry induced Berry curvature dipole (BCD) arising from asymmetric momentum space distribution of Berry curvature leads to second order nonlinear Hall effect (NHE). We observed strong bipolar distribution of Berry curvature. Consequently, BCD components attain maximum $D_{xx}=D_{yy}\approx0.169$ and $D_{zz}\approx-0.338$ at $E-E_F=0.06$~eV, corresponding to estimated nonlinear Hall current densities of $|j_x|\approx5.04\times10^{10}~\mathrm{A\,m^{-2}}$ and $|j_y|\approx2.52\times10^{10}~\mathrm{A\,m^{-2}}$. This reveal an anisotropic nonlinear Hall response. Furthermore, a distinct consequence of structural handedness emerges from the comparison of two enantiomers. The reversal of crystal chirality changes the signs of KW fermions and Berry curvature distributions, accompanied by opposite Fermi arc circulation. This reversal is also reflected in BCD and NHE. The handedness dependent reversal provides a possible electrical signature for distinguishing enantiomers and suggests opportunities for exploring chirality dependent nonlinear electronic responses.

The intrinsic spin Hall response provides a complementary perspective on the role of topological electronic states in transverse spin transport. Its pronounced maximum of $254.83 (\hbar/e)\mathrm{Scm^{-1}}$ at $E-E_F=-0.70$~eV occurs close to the triply charged KW point at $\Gamma$, suggesting that the energetic alignment of higher charge KW fermion, together with nearby spin Berry curvature hot spots, contributes to the enhanced response. This value exceeds the previously reported C40 compounds and  suggests an alternative pathway for amplifying the spin response, distinct from the conventional SOC driven anticrossing mechanism found in gapped systems. The coexistence of a substantial density of states at the Fermi level and ASOC splitting further raises the possibility of superconductivity in WAl$_2$. Taken together, these results show that WAl$_2$ brings multiple higher charge Kramers--Weyl phases, metallic fermiology, quantum oscillations, and Berry curvature driven charge and spin responses within a single material with potential applications in spintronics and nonlinear electronics. Our work motivates further exploration of multiple physical responses induced by quantum geometry. 

\section*{Acknowledgments}
SP (DST/INSPIRE Fellowship/2022/IF220593) highly appreciates the Department of Science and Technology (DST), Government of India, for the Inspire Fellowship. MS is thankful to Prof. Prafulla K. Jha for allowing to work on this problem as intern. This research used computational resources at PARAM Rudra 20 PF HPC System and 'PARAM Porul' at NIT Trichy provided by the Centre for Development of Advanced Computing (C-DAC) under the National Supercomputing Mission (NSM), India's flagship initiative jointly steered by the Ministry of Electronics and Information Technology (MeitY) and the Department of Science and Technology (DST). Authors are thankful to Anusandhan National Research Foundation (ANRF), Government of India, for financial assistance to develop computational resources through ANRF-PAIR project (Project No. ANRF/PAIR/2025/000017/PAIR). Some of the calculations were carried out at the Poznań Supercomputing and Networking Center (PSNC) under Grant No.~pl0331-03.
	
\section*{DATA AVAILABILITY}
The data that support the findings of this article are not publicly available. The data will be available from the authors upon reasonable request.
	

	\newpage
	
	\bibliographystyle{apsrev4-2}
	\bibliography{ref}

@article{Fang2012,
  title = {Multi-Weyl Topological Semimetals Stabilized by Point Group Symmetry},
  author = {Fang, Chen and Gilbert, Matthew J. and Dai, Xi and Bernevig, B. Andrei},
  journal = {Phys. Rev. Lett.},
  volume = {108},
  issue = {26},
  pages = {266802},
  numpages = {5},
  year = {2012},
  month = {Jun},
  publisher = {American Physical Society},
  doi = {10.1103/PhysRevLett.108.266802},
  url = {https://link.aps.org/doi/10.1103/PhysRevLett.108.266802}
}

@article{Tsutsumi2013,
author = {Tsutsumi, Hiroki and Miura, Yasunao and Harima, Hisatomo and Honda, Fuminori and Onuki, Yoshichika},
doi = {10.7566/JPSJ.82.094604},
issn = {00319015},
journal = {Journal of the Physical Society of Japan},
number = {9},
pages = {094604},
title = {{Split fermi surface properties of VSi2 without mirror planes in the chiral crystal structure}},
volume = {82},
year = {2013}
}

@article{Emmanouilidou2020,
  title = {Fermiology and type-I superconductivity in the chiral superconductor $\mathrm{Nb}{\mathrm{Ge}}_{2}$ with Kramers-Weyl fermions},
  author = {Emmanouilidou, Eve and Mardanya, Sougata and Xing, Jie and Reddy, P. V. Sreenivasa and Agarwal, Amit and Chang, Tay-Rong and Ni, Ni},
  journal = {Phys. Rev. B},
  volume = {102},
  issue = {23},
  pages = {235144},
  numpages = {9},
  year = {2020},
  month = {Dec},
  publisher = {American Physical Society},
  doi = {10.1103/PhysRevB.102.235144},
  url = {https://link.aps.org/doi/10.1103/PhysRevB.102.235144}
}

@article{Tsirkin2017,
author = {Tsirkin, Stepan S and Souza, Ivo and Vanderbilt, David},
doi = {10.1103/PhysRevB.96.045102},
issn = {24699969},
journal = {Physical Review B},
number = {4},
pages = {1--15},
title = {{Composite Weyl nodes stabilized by screw symmetry with and without time-reversal invariance}},
volume = {96},
year = {2017}
}

@article{Sato2025,
  author  = {Sato, Yoshiki J. and Aoki, Dai and Harima, Hisatomo},
  title   = {Recent advances in the study of chiral-lattice
             superconductor NbGe2 with topologically
             nontrivial electronic states},
  journal = {Journal of Physics: Condensed Matter},
  volume  = {37},
  number  = {50},
  pages   = {503001},
  year    = {2025},
  doi     = {10.1088/1361-648X/ae2225}
}

@article{Chang2018,
archivePrefix = {arXiv},
arxivId = {1611.07925},
author = {Chang, Guoqing and Wieder, Benjamin J and Schindler, Frank and Sanchez, Daniel S and Belopolski, Ilya and Huang, Shin Ming and Singh, Bahadur and Wu, Di and Chang, Tay Rong and Neupert, Titus and Xu, Su Yang and Lin, Hsin and Hasan, M. Zahid},
doi = {10.1038/s41563-018-0169-3},
eprint = {1611.07925},
issn = {14764660},
journal = {Nature Materials},
number = {11},
pages = {978--985},
pmid = {30275564},
title = {{Topological quantum properties of chiral crystals}},
volume = {17},
year = {2018}
}

@article{Dehghan2025,
doi = {10.1088/1361-648X/adeef0},
url = {https://doi.org/10.1088/1361-648X/adeef0},
year = {2025},
month = {jul},
publisher = {IOP Publishing},
volume = {37},
number = {30},
pages = {303002},
author = {Dehghan, Ali and Chen, Jiali and Jiang, Wei},
title = {Enhancing intrinsic spin Hall effect: insights into chiral crystals and topological materials},
journal = {Journal of Physics: Condensed Matter}
}

@article{Miyake2025,
author = {Miyake, Masataka and Tanaka, Natsumi and Nakamura, Naoki and Ōnuki, Yoshichika and Aoki, Dai and Nakamura, Ai and Harima, Hisatomo and Higashinaka, Ryuji and Aoki, Yuji and Matsuda, Tatsuma D.},
doi = {10.7566/JPSJ.94.034703},
issn = {13474073},
journal = {Journal of the Physical Society of Japan},
month = {mar},
number = {3},
title = {{de Haas–van Alphen Effect and Characteristic Fermi Surface Properties of PdGa with the Chiral Cubic Structure}},
url = {https://journals.jps.jp/doi/10.7566/JPSJ.94.034703},
volume = {94},
year = {2025}
}

@article{Alexandradinata2023,
   author = "Alexandradinata, A. and Glazman, Leonid",
   title = "Fermiology of Topological Metals", 
   journal= "Annual Review of Condensed Matter Physics",
   year = "2023",
   volume = "14",
   number = "Volume 14, 2023",
   pages = "261-309",
   doi = "https://doi.org/10.1146/annurev-conmatphys-040721-021331",
   url = "https://www.annualreviews.org/content/journals/10.1146/annurev-conmatphys-040721-021331",
   publisher = "Annual Reviews",
   issn = "1947-5462",
   type = "Journal Article",
  }

@article{Pu2026,
  title = {Enantiomeric pairs of chiral topological quasiparticles in chiral space groups},
  author = {Pu, Ruixi and Guan, Keqi and Wang, Houhao and Fan, Dongze and Huang, Liangliang and Wan, Xiangang and Tang, Feng},
  journal = {Phys. Rev. B},
  volume = {114},
  issue = {12},
  pages = {125113},
  numpages = {18},
  year = {2026},
  month = {Aug},
  publisher = {American Physical Society},
  doi = {10.1103/svtx-nrf8},
  url = {https://link.aps.org/doi/10.1103/svtx-nrf8}
}

@article{Kwan2020,
  title = {Quantum oscillations probe the Fermi surface topology of the nodal-line semimetal CaAgAs},
  author = {Kwan, Y. H. and Reiss, P. and Han, Y. and Bristow, M. and Prabhakaran, D. and Graf, D. and McCollam, A. and Parameswaran, S. A. and Coldea, A. I.},
  journal = {Phys. Rev. Res.},
  volume = {2},
  issue = {1},
  pages = {012055(R)},
  numpages = {5},
  year = {2020},
  month = {Mar},
  publisher = {American Physical Society},
  doi = {10.1103/PhysRevResearch.2.012055},
  url = {https://link.aps.org/doi/10.1103/PhysRevResearch.2.012055}
}

@article{Hasan2021,
  author  = {Hasan, M. Zahid and Chang, Guoqing and Belopolski, Ilya and Bian, Guang and Xu, Su-Yang and Yin, Jia-Xin},
  title   = {Weyl, Dirac and high-fold chiral fermions in topological quantum matter},
  journal = {Nature Reviews Materials},
  year    = {2021},
  volume  = {6},
  number  = {9},
  pages   = {784--803},
  doi     = {10.1038/s41578-021-00301-3},
  url     = {https://doi.org/10.1038/s41578-021-00301-3}
}

@article{Qiao2018,
  title = {Calculation of intrinsic spin Hall conductivity by Wannier interpolation},
  author = {Qiao, Junfeng and Zhou, Jiaqi and Yuan, Zhe and Zhao, Weisheng},
  journal = {Phys. Rev. B},
  volume = {98},
  issue = {21},
  pages = {214402},
  numpages = {10},
  year = {2018},
  month = {Dec},
  publisher = {American Physical Society},
  doi = {10.1103/PhysRevB.98.214402},
  url = {https://link.aps.org/doi/10.1103/PhysRevB.98.214402}
}

@article{Brasse2013,
archivePrefix = {arXiv},
arxivId = {1304.5994},
author = {Brasse, M and Chioncel, L and Kune{\v{s}}, J. and Bauer, A and Regnat, A and Blum, C. G.F. and Wurmehl, S and Pfleiderer, C and Wilde, M A and Grundler, D},
doi = {10.1103/PhysRevB.88.155138},
eprint = {1304.5994},
issn = {10980121},
journal = {Physical Review B - Condensed Matter and Materials Physics},
number = {15},
pages = {155138},
title = {{De Haas-van Alphen effect and Fermi surface properties of single-crystal CrB2}},
volume = {88},
year = {2013}
}

@article{Bradlyn2017,
archivePrefix = {arXiv},
arxivId = {1703.02050},
author = {Bradlyn, Barry and Elcoro, L and Cano, Jennifer and Vergniory, M G and Wang, Zhijun and Felser, C and Aroyo, M I and Bernevig, B Andrei},
doi = {10.1038/nature23268},
eprint = {1703.02050},
issn = {14764687},
journal = {Nature},
number = {7663},
pages = {298--305},
pmid = {28726818},
publisher = {Nature Publishing Group},
title = {{Topological quantum chemistry}},
url = {http://dx.doi.org/10.1038/nature23268},
volume = {547},
year = {2017}
}

@article{Sodemann2015,
  title = {Quantum Nonlinear Hall Effect Induced by Berry Curvature Dipole in Time-Reversal Invariant Materials},
  author = {Sodemann, Inti and Fu, Liang},
  journal = {Phys. Rev. Lett.},
  volume = {115},
  issue = {21},
  pages = {216806},
  numpages = {5},
  year = {2015},
  month = {Nov},
  publisher = {American Physical Society},
  doi = {10.1103/PhysRevLett.115.216806},
  url = {https://link.aps.org/doi/10.1103/PhysRevLett.115.216806}
}

@article{Ma2019,
archivePrefix = {arXiv},
arxivId = {1809.09279},
author = {Ma, Qiong and Xu, Su Yang and Shen, Huitao and MacNeill, David and Fatemi, Valla and Chang, Tay Rong and {Mier Valdivia}, Andr{\'{e}}s M. and Wu, Sanfeng and Du, Zongzheng and Hsu, Chuang Han and Fang, Shiang and Gibson, Quinn D. and Watanabe, Kenji and Taniguchi, Takashi and Cava, Robert J. and Kaxiras, Efthimios and Lu, Hai Zhou and Lin, Hsin and Fu, Liang and Gedik, Nuh and Jarillo-Herrero, Pablo},
doi = {10.1038/s41586-018-0807-6},
eprint = {1809.09279},
issn = {14764687},
journal = {Nature},
number = {7739},
pages = {337--342},
pmid = {30559379},
publisher = {Springer US},
title = {{Observation of the nonlinear Hall effect under time-reversal-symmetric conditions}},
url = {http://dx.doi.org/10.1038/s41586-018-0807-6},
volume = {565},
year = {2019}
}

@article{Shvetsov2019,
archivePrefix = {arXiv},
arxivId = {1902.02699},
author = {Shvetsov, O. O. and Esin, V. D. and Timonina, A. V. and Kolesnikov, N. N. and Deviatov, E. V.},
doi = {10.1134/S0021364019110018},
eprint = {1902.02699},
issn = {10906487},
journal = {JETP Letters},
number = {11},
pages = {715--721},
title = {{Nonlinear Hall Effect in Three-Dimensional Weyl and Dirac Semimetals}},
volume = {109},
year = {2019}
}

@article{Zeng2021,
archivePrefix = {arXiv},
arxivId = {2009.05043},
author = {Zeng, Chuanchang and Nandy, Snehasish and Tewari, Sumanta},
doi = {10.1103/PhysRevB.103.245119},
eprint = {2009.05043},
issn = {24699969},
journal = {Physical Review B},
number = {24},
pages = {245119},
publisher = {American Physical Society},
title = {{Nonlinear transport in Weyl semimetals induced by Berry curvature dipole}},
volume = {103},
year = {2021}
}

@article{Bandyopadhyay2022,
doi = {10.1088/2053-1583/ac6f63},
url = {https://doi.org/10.1088/2053-1583/ac6f63},
year = {2022},
month = {jun},
publisher = {IOP Publishing},
volume = {9},
number = {3},
pages = {035013},
author = {Bandyopadhyay, Arka and Joseph, Nesta Benno and Narayan, Awadhesh},
title = {Electrically switchable giant Berry curvature dipole in silicene, germanene and stanene},
journal = {2D Materials}
}

@article{Joseph2024,
archivePrefix = {arXiv},
arxivId = {2407.17775},
author = {Joseph, Nesta Benno and Bandyopadhyay, Arka and Narayan, Awadhesh},
doi = {10.1021/acs.chemmater.4c00427},
eprint = {2407.17775},
issn = {15205002},
journal = {Chemistry of Materials},
pages = {8602--8612},
publisher = {American Chemical Society},
title = {{Chirality-Tunable Nonlinear Hall Effect}},
volume = {36},
year = {2024}
}

@misc{Rani2026,
      title={Symmetry-Engineered Nonlinear Hall Response and Optical Response in Strained Monolayer Janus AsTeBr}, 
      author={Dimple Rani and Ritesh Ranjan Badhai and Gayatri Panda and Subrata Jana and Prasanjit Samal},
      year={2026},
      eprint={2607.17385},
      archivePrefix={arXiv},
      primaryClass={cond-mat.mtrl-sci},
      url={https://arxiv.org/abs/2607.17385}, 
}

@article{Sun2026,
author = {Sun, Hanbo and Wu, Chao and Dong, Pengqiang and She, Yanchao and Li, Ping},
doi = {10.1103/yqgb-td6q},
issn = {2469-9950},
journal = {Physical Review B},
month = {sep},
number = {16},
pages = {165407},
publisher = {American Physical Society},
title = {{Dual Hall effects in the altermagnet <math> <mrow> <msub> <mi>Ta</mi> <mn>2</mn> </msub> <msub> <mi>Te</mi> <mn>2</mn> </msub> <mi mathvariant="normal">O</mi> </mrow> </math> with strain-induced valley switching effect}},
url = {https://doi.org/10.1103/yqgb-td6q https://link.aps.org/doi/10.1103/yqgb-td6q},
volume = {114},
year = {2026}
}

@article{Zhou2026,
  title = {Nonvolatile control of nonlinear Hall and circular photogalvanic effects via Berry curvature dipole in multiferroic monolayer ${\mathrm{CrNBr}}_{2}$},
  author = {Zhou, Wenzhe and Li, Yinheng and Zhang, Dehe and Zheng, Guibo and Ouyang, Fangping},
  journal = {Phys. Rev. B},
  volume = {113},
  issue = {19},
  pages = {195126},
  numpages = {9},
  year = {2026},
  month = {May},
  publisher = {American Physical Society},
  doi = {10.1103/43bn-1d17},
  url = {https://link.aps.org/doi/10.1103/43bn-1d17}
}

@article{Liao2026,
author = {Liao, Zishan and Xu, Zhiming and Wang, Erqing and Huang, H.},
doi = {10.1103/9wgk-g3dd},
issn = {2469-9950},
journal = {Physical Review B},
month = {apr},
number = {16},
pages = {165103},
title = {{Strain-tunable Berry curvature dipole and doping-enhanced shift current in two-dimensional Janus Bi 2 TeSe 2}},
url = {https://link.aps.org/doi/10.1103/9wgk-g3dd},
volume = {113},
year = {2026}
}

@article{Gao2026QG,
  title = {Quantum geometry phenomena in condensed matter systems},
  author = {Gao, Anyuan and Nagaosa, Naoto and Ni, Ni and Xu, Su-Yang},
  journal = {Rev. Mod. Phys.},
  volume = {98},
  issue = {3},
  pages = {035005},
  numpages = {67},
  year = {2026},
  month = {Sep},
  publisher = {American Physical Society},
  doi = {10.1103/f78t-ky69},
  url = {https://link.aps.org/doi/10.1103/f78t-ky69}
}

@article{Liu2025QG,
    author = {Liu, Tianyu and Qiang, Xiao-Bin and Lu, Hai-Zhou and Xie, X C},
    title = {Quantum geometry in condensed matter},
    journal = {National Science Review},
    volume = {12},
    number = {3},
    pages = {nwae334},
    year = {2025},
    month = {03},
    issn = {2095-5138},
    doi = {10.1093/nsr/nwae334},
    url = {https://doi.org/10.1093/nsr/nwae334},
    eprint = {https://academic.oup.com/nsr/article-pdf/12/3/nwae334/59204701/nwae334.pdf},
}

@article{Berry1984,
    author = {Berry, Michael Victor},
    title = {Quantal phase factors accompanying adiabatic changes},
    journal = {Proceedings of the Royal Society of London. A. Mathematical and Physical Sciences},
    volume = {392},
    number = {1802},
    pages = {45-57},
    year = {1984},
    month = {03},
    issn = {0080-4630},
    doi = {10.1098/rspa.1984.0023},
    url = {https://doi.org/10.1098/rspa.1984.0023},
    eprint = {https://royalsocietypublishing.org/rspa/article-pdf/392/1802/45/65511/rspa.1984.0023.pdf},
}

@article{Xiao2010,
  title = {{Berry phase effects on electronic properties}},
  author = {Xiao, Di and Chang, Ming-Che and Niu, Qian},
  journal = {Rev. Mod. Phys.},
  volume = {82},
  issue = {3},
  pages = {1959--2007},
  numpages = {0},
  year = {2010},
  month = {Jul},
  publisher = {American Physical Society},
  doi = {10.1103/RevModPhys.82.1959},
  url = {https://link.aps.org/doi/10.1103/RevModPhys.82.1959}
}

@article{Armitage2018,
  title = {{Weyl and Dirac semimetals in three-dimensional solids}},
  author = {Armitage, N. P. and Mele, E. J. and Vishwanath, Ashvin},
  journal = {Rev. Mod. Phys.},
  volume = {90},
  issue = {1},
  pages = {015001},
  numpages = {57},
  year = {2018},
  month = {Jan},
  publisher = {American Physical Society},
  doi = {10.1103/RevModPhys.90.015001},
  url = {https://link.aps.org/doi/10.1103/RevModPhys.90.015001}
}

@article{Kousaka2023,
author = {Kousaka, Yusuke and Sayo, Taisei and Iwasaki, Satoshi and Saki, Ryo and Shimada, Chiho and Shishido, Hiroaki and Togawa, Yoshihiko},
doi = {10.35848/1347-4065/aca8e2},
issn = {0021-4922},
journal = {Japanese Journal of Applied Physics},
month = {jan},
number = {1},
pages = {015506},
publisher = {IOP Publishing},
title = {{Chirality-selected crystal growth and spin polarization over centimeters of transition metal disilicide crystals}},
url = {https://iopscience.iop.org/article/10.35848/1347-4065/aca8e2},
volume = {62},
year = {2023}
}

@article{Zhang2024Chirality,
author = {Zhang, Yudi and Ma, Yuzhe and Sun, Wen and Li, Wei and Li, Guowei},
title = {Structural and Electronic Chirality in Inorganic Crystals: from Construction to Application},
journal = {Chemistry – A European Journal},
volume = {30},
number = {34},
pages = {e202400436},
doi = {https://doi.org/10.1002/chem.202400436},
url = {https://chemistry-europe.onlinelibrary.wiley.com/doi/abs/10.1002/chem.202400436},
eprint = {https://chemistry-europe.onlinelibrary.wiley.com/doi/pdf/10.1002/chem.202400436},
year = {2024}
}

@article{Sanchez2019,
author = {Sanchez, Daniel S. and Belopolski, Ilya and Cochran, Tyler A. and Xu, Xitong and Yin, Jia Xin and Chang, Guoqing and Xie, Weiwei and Manna, Kaustuv and S{\"{u}}{\ss}, Vicky and Huang, Cheng Yi and Alidoust, Nasser and Multer, Daniel and Zhang, Songtian S. and Shumiya, Nana and Wang, Xirui and Wang, Guang Qiang and Chang, Tay Rong and Felser, Claudia and Xu, Su Yang and Jia, Shuang and Lin, Hsin and Hasan, M. Zahid},
doi = {10.1038/s41586-019-1037-2},
issn = {14764687},
journal = {Nature},
number = {7749},
pages = {500--505},
pmid = {30894753},
publisher = {Springer US},
title = {{Topological chiral crystals with helicoid-arc quantum states}},
url = {http://dx.doi.org/10.1038/s41586-019-1037-2},
volume = {567},
year = {2019}
}

@article{Rao2019,
author = {Rao, Zhicheng and Li, Hang and Zhang, Tiantian and Tian, Shangjie and Li, Chenghe and Fu, Binbin and Tang, Cenyao and Wang, Le and Li, Zhilin and Fan, Wenhui and Li, Jiajun and Huang, Yaobo and Liu, Zhehong and Long, Youwen and Fang, Chen and Weng, Hongming and Shi, Youguo and Lei, Hechang and Sun, Yujie and Qian, Tian and Ding, Hong},
doi = {10.1038/s41586-019-1031-8},
issn = {0028-0836},
journal = {Nature},
month = {mar},
number = {7749},
pages = {496--499},
pmid = {30894751},
publisher = {Springer US},
title = {{Observation of unconventional chiral fermions with long Fermi arcs in CoSi}},
url = {http://dx.doi.org/10.1038/s41586-019-1031-8 https://www.nature.com/articles/s41586-019-1031-8},
volume = {567},
year = {2019}
}

@article{Sharma2026,
author = {Sharma, Avdhesh K. and Chatterjee, Snehashish and Yanda, Premakumar and Felser, Claudia and Shekhar, Chandra},
doi = {10.1038/s41535-026-00902-7},
issn = {2397-4648},
journal = {npj Quantum Materials},
month = {jun},
number = {1},
pages = {55},
publisher = {Springer US},
title = {{Topological quantum materials: kagome, chiral, and square-net frameworks}},
url = {https://www.nature.com/articles/s41535-026-00902-7},
volume = {11},
year = {2026}
}

@article{Smidman2017supercondutivity,
doi = {10.1088/1361-6633/80/3/036501},
url = {https://doi.org/10.1088/1361-6633/80/3/036501},
year = {2017},
month = {jan},
publisher = {IOP Publishing},
volume = {80},
number = {3},
pages = {036501},
author = {Smidman, M and Salamon, M B and Yuan, H Q and Agterberg, D F},
title = {Superconductivity and spin–orbit coupling in non-centrosymmetric materials: a review},
journal = {Reports on Progress in Physics}
}

@article{Sinova2015,
  title = {Spin Hall effects},
  author = {Sinova, Jairo and Valenzuela, Sergio O. and Wunderlich, J. and Back, C. H. and Jungwirth, T.},
  journal = {Rev. Mod. Phys.},
  volume = {87},
  issue = {4},
  pages = {1213--1260},
  numpages = {47},
  year = {2015},
  month = {Oct},
  publisher = {American Physical Society},
  doi = {10.1103/RevModPhys.87.1213},
  url = {https://link.aps.org/doi/10.1103/RevModPhys.87.1213}
}

@article{Du2018BCD_WTe2,
  title = {Band Signatures for Strong Nonlinear Hall Effect in Bilayer ${\mathrm{WTe}}_{2}$},
  author = {Du, Z. Z. and Wang, C. M. and Lu, Hai-Zhou and Xie, X. C.},
  journal = {Phys. Rev. Lett.},
  volume = {121},
  issue = {26},
  pages = {266601},
  numpages = {6},
  year = {2018},
  month = {Dec},
  publisher = {American Physical Society},
  doi = {10.1103/PhysRevLett.121.266601},
  url = {https://link.aps.org/doi/10.1103/PhysRevLett.121.266601}
}

@article{Xiao2020_BCDLiOsO3,
  title = {Electrical detection of ferroelectriclike metals through the nonlinear Hall effect},
  author = {Xiao, Rui-Chun and Shao, Ding-Fu and Huang, Wenjuan and Jiang, Hua},
  journal = {Phys. Rev. B},
  volume = {102},
  issue = {2},
  pages = {024109},
  numpages = {7},
  year = {2020},
  month = {Jul},
  publisher = {American Physical Society},
  doi = {10.1103/PhysRevB.102.024109},
  url = {https://link.aps.org/doi/10.1103/PhysRevB.102.024109}
}

@article{Pang2024TaAs,
  title = {Tuning of Berry-curvature dipole in TaAs slabs: An effective route to enhance the nonlinear Hall response},
  author = {Pang, Hongsheng and Jin, Gan and He, Lixin},
  journal = {Phys. Rev. Mater.},
  volume = {8},
  issue = {4},
  pages = {043403},
  numpages = {6},
  year = {2024},
  month = {Apr},
  publisher = {American Physical Society},
  doi = {10.1103/PhysRevMaterials.8.043403},
  url = {https://link.aps.org/doi/10.1103/PhysRevMaterials.8.043403}
}

@article{Lee2024NbIrTe4,
author = {Lee, Ji-Eun and Wang, Aifeng and Chen, Shuzhang and Kwon, Minseong and Hwang, Jinwoong and Cho, Minhyun and Son, Ki-Hoon and Han, Dong-Soo and Choi, Jun Woo and Kim, Young Duck and Mo, Sung-Kwan and Petrovic, Cedomir and Hwang, Choongyu and Park, Se Young and Jang, Chaun and Ryu, Hyejin},
doi = {10.1038/s41467-024-47643-4},
issn = {2041-1723},
journal = {Nature Communications},
month = {may},
number = {1},
pages = {3971},
pmid = {38729931},
publisher = {Springer US},
title = {{Spin-orbit-splitting-driven nonlinear Hall effect in NbIrTe4}},
url = {https://www.nature.com/articles/s41467-024-47643-4},
volume = {15},
year = {2024}
}

@article{Wu2025TaS2,
author = {Wu, Shengqiang and Tian, Wanghao and Li, Runlai and Han, Ziyi and Zhou, Xuan and Huang, Song and Li, Ping and Song, Peng and Zhao, Xiaoxu},
doi = {10.1016/j.matt.2025.102153},
issn = {25902385},
journal = {Matter},
month = {sep},
number = {9},
pages = {102153},
publisher = {Elsevier Inc.},
title = {{Self-intercalated 6R-TaS2 with reduced symmetry for room temperature nonlinear Hall effect}},
url = {https://doi.org/10.1016/j.matt.2025.102153 https://linkinghub.elsevier.com/retrieve/pii/S2590238525001961},
volume = {8},
year = {2025}
}

@article{Montiel2023spin_injectors,
  title = {Spin current injection via equal-spin Cooper pairs in ferromagnet/superconductor heterostructures},
  author = {Montiel, X. and Eschrig, M.},
  journal = {Phys. Rev. B},
  volume = {107},
  issue = {9},
  pages = {094513},
  numpages = {9},
  year = {2023},
  month = {Mar},
  publisher = {American Physical Society},
  doi = {10.1103/PhysRevB.107.094513},
  url = {https://link.aps.org/doi/10.1103/PhysRevB.107.094513}
}

@article{Guo2023_spin_transistors,
author = {Guo, Lidan and Hu, Shunhua and Gu, Xianrong and Zhang, Rui and Wang, Kai and Yan, Wenjing and Sun, Xiangnan},
title = {Emerging Spintronic Materials and Functionalities},
journal = {Advanced Materials},
volume = {36},
number = {22},
pages = {2301854},
doi = {https://doi.org/10.1002/adma.202301854},
url = {https://advanced.onlinelibrary.wiley.com/doi/abs/10.1002/adma.202301854},
eprint = {https://advanced.onlinelibrary.wiley.com/doi/pdf/10.1002/adma.202301854},
year = {2024}
}

@article{Amiri2024,
   author = "Khalili Amiri, Pedram and Phatak, Charudatta and Finocchio, Giovanni",
   title = "Prospects for Antiferromagnetic Spintronic Devices", 
   journal= "Annual Review of Materials Research",
   year = "2024",
   volume = "54",
   number = "Volume 54, 2024",
   pages = "117-142",
   doi = "https://doi.org/10.1146/annurev-matsci-080222-030535",
   url = "https://www.annualreviews.org/content/journals/10.1146/annurev-matsci-080222-030535",
   publisher = "Annual Reviews",
   issn = "1545-4118",
   type = "Journal Article",
  }

@article{VASPKresse1996,
author = {Kresse, G. and Furthm{\"{u}}ller, J.},
doi = {10.1016/0927-0256(96)00008-0},
issn = {09270256},
journal = {Computational Materials Science},
month = {jul},
number = {1},
pages = {15--50},
title = {{Efficiency of ab-initio total energy calculations for metals and semiconductors using a plane-wave basis set}},
url = {https://linkinghub.elsevier.com/retrieve/pii/0927025696000080},
volume = {6},
year = {1996}
}

@article{PAWBlochl1994,
author = {Bl{\"{o}}chl, P. E.},
doi = {10.1103/PhysRevB.50.17953},
issn = {0163-1829},
journal = {Physical Review B},
month = {dec},
number = {24},
pages = {17953--17979},
pmid = {9976227},
title = {{Projector augmented-wave method}},
url = {https://link.aps.org/doi/10.1103/PhysRevB.50.17953},
volume = {50},
year = {1994}
}

@article{GGA,
  title = {Generalized Gradient Approximation Made Simple},
  author = {Perdew, John P. and Burke, Kieron and Ernzerhof, Matthias},
  journal = {Phys. Rev. Lett.},
  volume = {77},
  issue = {18},
  pages = {3865--3868},
  numpages = {0},
  year = {1996},
  month = {Oct},
  publisher = {American Physical Society},
  doi = {10.1103/PhysRevLett.77.3865},
  url = {https://link.aps.org/doi/10.1103/PhysRevLett.77.3865}
}

@article{WANNIER90Mostofi2014,
author = {Mostofi, Arash A. and Yates, Jonathan R. and Pizzi, Giovanni and Lee, Young-Su and Souza, Ivo and Vanderbilt, David and Marzari, Nicola},
doi = {10.1016/j.cpc.2014.05.003},
issn = {00104655},
journal = {Computer Physics Communications},
month = {aug},
number = {8},
pages = {2309--2310},
title = {{An updated version of wannier90: A tool for obtaining maximally-localised Wannier functions}},
url = {https://linkinghub.elsevier.com/retrieve/pii/S001046551400157X},
volume = {185},
year = {2014}
}

@article{WANNIERTOOLS2018,
author = {Wu, QuanSheng and Zhang, ShengNan and Song, Hai-Feng and Troyer, Matthias and Soluyanov, Alexey A.},
doi = {10.1016/j.cpc.2017.09.033},
issn = {00104655},
journal = {Computer Physics Communications},
month = {mar},
pages = {405--416},
title = {{WannierTools: An open-source software package for novel topological materials}},
url = {https://linkinghub.elsevier.com/retrieve/pii/S0010465517303442},
volume = {224},
year = {2018}
}

@article{Saurav2025,
  title = {Composite quantum materials hosting giant Rashba effect, topological phases, and electric polarization control},
  author = {Patel, Saurav and Jha, Prafulla K.},
  journal = {Phys. Rev. B},
  volume = {112},
  issue = {7},
  pages = {075131},
  numpages = {12},
  year = {2025},
  month = {Aug},
  publisher = {American Physical Society},
  doi = {10.1103/6ly5-112g},
  url = {https://link.aps.org/doi/10.1103/6ly5-112g}
}

@article{Jay2025APL,
  author = {Panchal, Jay and Patel, Saurav and Jha, Prafulla K.},
  title = {First-principles prediction of symmetry-enforced multifold fermionic and bosonic degeneracies in {M3Sn} ({M = V, Nb}) compounds},
  journal = {Applied Physics Letters},
  volume = {127},
  number = {14},
  pages = {143102},
  year = {2025},
  doi = {10.1063/5.0288878},
  url = {https://doi.org/10.1063/5.0288878}
}

@book{Cambridge_Univ_Press_1984,
author = {Shoenberg, D.},
doi = {10.1017/CBO9780511897870},
isbn = {9780521224802},
month = {jan},
publisher = {Cambridge University Press},
title = {{Magnetic Oscillations in Metals}},
url = {https://www.cambridge.org/core/product/identifier/9780511897870/type/book},
year = {1984}
}

@article{Saini_YSi_2021,
  title = {Fermi surface studies of the topologically nontrivial compound YSi},
  author = {Saini, Vikas and Sasmal, Souvik and Kulkarni, Ruta and Singh, Bahadur and Thamizhavel, A.},
  journal = {Phys. Rev. B},
  volume = {103},
  issue = {19},
  pages = {195154},
  numpages = {7},
  year = {2021},
  month = {May},
  publisher = {American Physical Society},
  doi = {10.1103/PhysRevB.103.195154},
  url = {https://link.aps.org/doi/10.1103/PhysRevB.103.195154}
}

@article{Saini_PtAl_2022,
  title = {Analysis of unconventional chiral fermions in a noncentrosymmetric chiral crystal PtAl},
  author = {Saini, Vikas and Sasmal, Souvik and Kulkarni, Ruta and Singh, Bahadur and Thamizhavel, A. and Nakamura, Ai and Aoki, Dai},
  journal = {Phys. Rev. B},
  volume = {106},
  issue = {12},
  pages = {125126},
  numpages = {8},
  year = {2022},
  month = {Sep},
  publisher = {American Physical Society},
  doi = {10.1103/PhysRevB.106.125126},
  url = {https://link.aps.org/doi/10.1103/PhysRevB.106.125126}
}

@article{WAL2_2008_synthesis,
title = {Crystal structure and high-pressure studies of WAl2, an aluminide crystallizing with the CrSi2 structure type},
journal = {Journal of Solid State Chemistry},
volume = {181},
number = {10},
pages = {2719-2724},
year = {2008},
issn = {0022-4596},
doi = {https://doi.org/10.1016/j.jssc.2008.06.053},
url = {https://www.sciencedirect.com/science/article/pii/S0022459608003472},
author = {Q.F. Gu and D.Y. Jung and G. Krauss and W. Steurer}
}

@article{Momma2011,
author = {Momma, Koichi and Izumi, Fujio},
doi = {10.1107/S0021889811038970},
issn = {0021-8898},
journal = {Journal of Applied Crystallography},
month = {dec},
number = {6},
pages = {1272--1276},
title = {{VESTA 3 for three-dimensional visualization of crystal, volumetric and morphology data}},
url = {https://scripts.iucr.org/cgi-bin/paper?S0021889811038970 https://journals.iucr.org/paper?S0021889811038970},
volume = {44},
year = {2011}
}

@article{Onsager_1952,
author = {L. Onsager},
title = {{Interpretation of the de Haas-van Alphen effect}},
journal = {The London, Edinburgh, and Dublin Philosophical Magazine and Journal of Science},
volume = {43},
number = {344},
pages = {1006--1008},
year = {1952},
publisher = {Taylor \& Francis},
doi = {10.1080/14786440908521019},
URL = {https://doi.org/10.1080/14786440908521019},
eprint = {https://doi.org/10.1080/14786440908521019}
}

@article{SKEAF_2012,
title = {Numerical extraction of de Haas–van Alphen frequencies from calculated band energies},
journal = {Computer Physics Communications},
volume = {183},
number = {2},
pages = {324-332},
year = {2012},
issn = {0010-4655},
doi = {https://doi.org/10.1016/j.cpc.2011.10.015},
url = {https://www.sciencedirect.com/science/article/pii/S001046551100347X},
author = {P.M.C. Rourke and S.R. Julian}
}

@article{WannierBerri_2021,
author = {Tsirkin, Stepan S},
doi = {10.1038/s41524-021-00498-5},
issn = {2057-3960},
journal = {npj Computational Materials},
month = {feb},
number = {1},
pages = {33},
publisher = {Springer US},
title = {{High performance Wannier interpolation of Berry curvature and related quantities with WannierBerri code}},
url = {http://dx.doi.org/10.1038/s41524-021-00498-5 https://www.nature.com/articles/s41524-021-00498-5},
volume = {7},
year = {2021}
}

@book{Ziman2001,
  author    = {J. M. Ziman},
  title     = {Electrons and Phonons: The Theory of Transport Phenomena in Solids},
  publisher = {Oxford University Press},
  year      = {2001}
}

@article{Sinova_SHC_2004,
  title = {Universal Intrinsic Spin Hall Effect},
  author = {Sinova, Jairo and Culcer, Dimitrie and Niu, Q. and Sinitsyn, N. A. and Jungwirth, T. and MacDonald, A. H.},
  journal = {Phys. Rev. Lett.},
  volume = {92},
  issue = {12},
  pages = {126603},
  numpages = {4},
  year = {2004},
  month = {Mar},
  publisher = {American Physical Society},
  doi = {10.1103/PhysRevLett.92.126603},
  url = {https://link.aps.org/doi/10.1103/PhysRevLett.92.126603}
}

@article{2005aCTA,
title = {Enantiomorph identification of transition-metal disilicides with the C40 structure (the space group of P6222 and P6422) by new convergent-beam electron diffraction method},
journal = {Acta Materialia},
volume = {53},
number = {1},
pages = {41-52},
year = {2005},
issn = {1359-6454},
doi = {https://doi.org/10.1016/j.actamat.2004.08.001},
url = {https://www.sciencedirect.com/science/article/pii/S1359645404004689},
author = {Hiroki Sakamoto and Akihiro Fujii and Katsushi Tanaka and Haruyuki Inui}
}

@article{Tanaka2001,
author = {Tanaka, K and Nawata, K and Inui, H and Yamaguchi, M and Koiwa, M},
doi = {10.1016/S0966-9795(01)00048-6},
issn = {09669795},
journal = {Intermetallics},
month = {jul},
number = {7},
pages = {603--607},
title = {{Refinement of crystallographic parameters in transition metal disilicides with the C11b, C40 and C54 structures}},
url = {https://linkinghub.elsevier.com/retrieve/pii/S0966979501000486},
volume = {9},
year = {2001}
}

@misc{supplemental_material,
  note = {See Supplemental Material for AIMD simulations,
          orbital projections, comparison of DFT and
          Wannier90 band structures, density of states,
          dHvA frequency splitting, and symmetry analysis
          of Kramers--Weyl nodes},
  year = {2026}
}

@article{Yang2026ScSi2,
  title   = {Emergent topology and superconductivity in hexagonal
             {ScSi2}: A first-principles perspective},
  author  = {Saoula, B. and Menad, A. and Beldi, L. and
             Ferhat, M. and Bouhafs, B. and Kanoun, M. B. and
             Goumri-Said, S.},
  journal = {Journal of Physics and Chemistry of Solids},
  volume  = {215},
  pages   = {113708},
  year    = {2026},
  doi     = {10.1016/j.jpcs.2026.113708}
}

@article{2020_natcomm_PdGa,
author = {Sessi, Paolo and Fan, Feng-Ren and K{\"u}ster, Felix and Manna, Kaustuv and Schr{\"o}ter, Niels B. M. and Ji, Jing-Rong and Stolz, Samuel and Krieger, Jonas A. and Pei, Ding and Kim, Timur K. and Dudin, Pavel and Cacho, Cephise and Widmer, Roland and Borrmann, Horst and Shi, Wujun and Chang, Kai and Sun, Yan and Felser, Claudia and Parkin, Stuart S. P.},
doi = {10.1038/s41467-020-17261-x},
issn = {2041-1723},
journal = {Nature Communications},
month = {jul},
number = {1},
pages = {3507},
title = {{Handedness-dependent quasiparticle interference in the two enantiomers of the topological chiral semimetal PdGa}},
url = {https://www.nature.com/articles/s41467-020-17261-x},
volume = {11},
year = {2020}
}

@article{Li2019_CoSi,
author = {Li, Hang and Xu, Sheng and Rao, Zhi-Cheng and Zhou, Li-Qin and Wang, Zhi-Jun and Zhou, Shi-Ming and Tian, Shang-Jie and Gao, Shun-Ye and Li, Jia-Jun and Huang, Yao-Bo and Lei, He-Chang and Weng, Hong-Ming and Sun, Yu-Jie and Xia, Tian-Long and Qian, Tian and Ding, Hong},
doi = {10.1038/s41467-019-13435-4},
issn = {2041-1723},
journal = {Nature Communications},
month = {dec},
number = {1},
pages = {5505},
title = {{Chiral fermion reversal in chiral crystals}},
url = {https://www.nature.com/articles/s41467-019-13435-4},
volume = {10},
year = {2019}
}

@article{Yu2022Encyclopedia,
author = {Yu, Zhi-Ming and Zhang, Zeying and Liu, Gui-Bin and Wu, Weikang and Li, Xiao-Ping and Zhang, Run-Wu and Yang, Shengyuan A. and Yao, Yugui},
doi = {10.1016/j.scib.2021.10.023},
issn = {20959273},
journal = {Science Bulletin},
month = {feb},
number = {4},
pages = {375--380},
title = {{Encyclopedia of emergent particles in three-dimensional crystals}},
url = {https://linkinghub.elsevier.com/retrieve/pii/S2095927321006927},
volume = {67},
year = {2022}
}

@article{Muller_2020,
  title = {Determination of the Fermi surface and field-induced quasiparticle tunneling around the Dirac nodal loop in ZrSiS},
  author = {M{\"u}ller, C. S. A. and Khouri, T. and van Delft, M. R. and Pezzini, S. and Hsu, Y.-T. and Ayres, J. and Breitkreiz, M. and Schoop, L. M. and Carrington, A. and Hussey, N. E. and Wiedmann, S.},
  journal = {Phys. Rev. Res.},
  volume = {2},
  issue = {2},
  pages = {023217},
  numpages = {13},
  year = {2020},
  month = {May},
  publisher = {American Physical Society},
  doi = {10.1103/PhysRevResearch.2.023217},
  url = {https://link.aps.org/doi/10.1103/PhysRevResearch.2.023217}
}

@article{Nguyen_2022,
  title = {Fermiology of the Dirac type-II semimetal candidates (Ni,Zr)${\mathrm{Te}}_{2}$ using de Haas--van Alphen oscillations},
  author = {Nguyen, T. and Aryal, N. and Pokharel, Bal K. and Harnagea, L. and Mierstchin, D. and Popovi{\ifmmode \acute{c}\else \'{c}\fi{}}, Dragana and Graf, D. E. and Shrestha, K.},
  journal = {Phys. Rev. B},
  volume = {106},
  issue = {7},
  pages = {075154},
  numpages = {9},
  year = {2022},
  month = {Aug},
  publisher = {American Physical Society},
  doi = {10.1103/PhysRevB.106.075154},
  url = {https://link.aps.org/doi/10.1103/PhysRevB.106.075154}
}

@article{Wu_2025_RuO2,
  title = {Fermi Surface of ${\mathrm{RuO}}_{2}$ Measured by Quantum Oscillations},
  author = {Wu, Zheyu and Long, Mengmeng and Chen, Hanyi and Paul, Shubhankar and Matsuki, Hisakazu and Zheliuk, Oleksandr and Zeitler, Uli and Li, Gang and Zhou, Rui and Zhu, Zengwei and Graf, Dave and Weinberger, Theodore I. and Grosche, F. Malte and Maeno, Yoshiteru and Eaton, Alexander G.},
  journal = {Phys. Rev. X},
  volume = {15},
  issue = {3},
  pages = {031044},
  numpages = {13},
  year = {2025},
  month = {Aug},
  publisher = {American Physical Society},
  doi = {10.1103/5js8-2hj8},
  url = {https://link.aps.org/doi/10.1103/5js8-2hj8}
}

@article{Matsyshyn2019,
  title = {Nonlinear Hall Acceleration and the Quantum Rectification Sum Rule},
  author = {Matsyshyn, O. and Sodemann, I.},
  journal = {Phys. Rev. Lett.},
  volume = {123},
  issue = {24},
  pages = {246602},
  numpages = {6},
  year = {2019},
  month = {Dec},
  publisher = {American Physical Society},
  doi = {10.1103/PhysRevLett.123.246602},
  url = {https://link.aps.org/doi/10.1103/PhysRevLett.123.246602}
}

@article{Arka2024,
title = {Non-linear Hall effects: Mechanisms and materials},
journal = {Materials Today Electronics},
volume = {8},
pages = {100101},
year = {2024},
issn = {2772-9494},
doi = {https://doi.org/10.1016/j.mtelec.2024.100101},
url = {https://www.sciencedirect.com/science/article/pii/S2772949424000135},
author = {Arka Bandyopadhyay and Nesta Benno Joseph and Awadhesh Narayan}
}

@article{Kang2019,
author = {Kang, Kaifei and Li, Tingxin and Sohn, Egon and Shan, Jie and Mak, Kin Fai},
doi = {10.1038/s41563-019-0294-7},
issn = {1476-1122},
journal = {Nature Materials},
month = {apr},
number = {4},
pages = {324--328},
title = {{Nonlinear anomalous Hall effect in few-layer WTe2}},
url = {https://www.nature.com/articles/s41563-019-0294-7},
volume = {18},
year = {2019}
}

@article{Roy2022,
doi = {10.1088/1361-648X/ac8091},
url = {https://doi.org/10.1088/1361-648X/ac8091},
year = {2022},
month = {jul},
publisher = {IOP Publishing},
volume = {34},
number = {38},
pages = {385301},
author = {Roy, Saswata and Narayan, Awadhesh},
title = {Non-linear Hall effect in multi-Weyl semimetals},
journal = {Journal of Physics: Condensed Matter}
}

@article{Saha2023,
doi = {10.1088/1361-648X/acf1eb},
url = {https://doi.org/10.1088/1361-648X/acf1eb},
year = {2023},
month = {sep},
publisher = {IOP Publishing},
volume = {35},
number = {48},
pages = {485301},
author = {Saha, Soumadeep and Narayan, Awadhesh},
title = {Nonlinear Hall effect in Rashba systems with hexagonal warping},
journal = {Journal of Physics: Condensed Matter}
}

@article{Zhang2021_THz_NHE,
doi = {10.1073/pnas.2100736118},
author = {Yang Zhang  and Liang Fu },
title = {Terahertz detection based on nonlinear Hall effect without magnetic field},
journal = {Proceedings of the National Academy of Sciences},
volume = {118},
number = {21},
pages = {e2100736118},
year = {2021},
URL = {https://www.pnas.org/doi/abs/10.1073/pnas.2100736118},
eprint = {https://www.pnas.org/doi/pdf/10.1073/pnas.2100736118}
}

@book{Ramsden2011,
  author    = {Edward Ramsden},
  title     = {Hall-Effect Sensors: Theory and Application},
  publisher = {Elsevier},
  year      = {2011}
}

@article{Xiao2020,
  title = {Electrical detection of ferroelectriclike metals through the nonlinear Hall effect},
  author = {Xiao, Rui-Chun and Shao, Ding-Fu and Huang, Wenjuan and Jiang, Hua},
  journal = {Phys. Rev. B},
  volume = {102},
  issue = {2},
  pages = {024109},
  numpages = {7},
  year = {2020},
  month = {Jul},
  publisher = {American Physical Society},
  doi = {10.1103/PhysRevB.102.024109},
  url = {https://link.aps.org/doi/10.1103/PhysRevB.102.024109}
}

@article{Facio2018,
  title = {Strongly Enhanced Berry Dipole at Topological Phase Transitions in BiTeI},
  author = {Facio, Jorge I. and Efremov, Dmitri and Koepernik, Klaus and You, Jhih-Shih and Sodemann, Inti and van den Brink, Jeroen},
  journal = {Phys. Rev. Lett.},
  volume = {121},
  issue = {24},
  pages = {246403},
  numpages = {6},
  year = {2018},
  month = {Dec},
  publisher = {American Physical Society},
  doi = {10.1103/PhysRevLett.121.246403},
  url = {https://link.aps.org/doi/10.1103/PhysRevLett.121.246403}
}

@article{Sinha2022,
author = {Sinha, Subhajit and Adak, Pratap Chandra and Chakraborty, Atasi and Das, Kamal and Debnath, Koyendrila and Sangani, L. D. Varma and Watanabe, Kenji and Taniguchi, Takashi and Waghmare, Umesh V. and Agarwal, Amit and Deshmukh, Mandar M.},
doi = {10.1038/s41567-022-01606-y},
issn = {1745-2473},
journal = {Nature Physics},
month = {jul},
number = {7},
pages = {765--770},
title = {{Berry curvature dipole senses topological transition in a moir{\'{e}} superlattice}},
url = {https://www.nature.com/articles/s41567-022-01606-y},
volume = {18},
year = {2022}
}

@article{Zhang2022_ferro_bcd,
author = {Zhang, Cheng-Long and Liang, Tian and Kaneko, Yoshio and Nagaosa, Naoto and Tokura, Yoshinori},
doi = {10.1038/s41535-022-00512-z},
issn = {2397-4648},
journal = {npj Quantum Materials},
month = {oct},
number = {1},
pages = {103},
title = {{Giant Berry curvature dipole density in a ferroelectric Weyl semimetal}},
url = {https://www.nature.com/articles/s41535-022-00512-z},
volume = {7},
year = {2022}
}

@article{Nishijima2023,
    author = {Nishijima, Taiki and Watanabe, Takuto and Sekiguchi, Hiroaki and Ando, Yuichiro and Shigematsu, Ei and Ohshima, Ryo and Kuroda, Shinji and Shiraishi, Masashi},
    title = {Ferroic Berry
Curvature Dipole in a Topological Crystalline
Insulator at Room Temperature},
    journal = {Nano Letters},
    volume = {23},
    number = {6},
    pages = {2247-2252},
    year = {2023},
    month = {03},
    issn = {1530-6984},
    doi = {10.1021/acs.nanolett.2c04900},
    url = {https://doi.org/10.1021/acs.nanolett.2c04900},
    eprint = {https://pubs.acs.org/nalefd/article-pdf/23/6/2247/2635467/nl2c04900.pdf},
}

@article{Kumar2021,
author = {Kumar, Dushyant and Hsu, Chuang-Han and Sharma, Raghav and Chang, Tay-Rong and Yu, Peng and Wang, Junyong and Eda, Goki and Liang, Gengchiau and Yang, Hyunsoo},
doi = {10.1038/s41565-020-00839-3},
issn = {1748-3387},
journal = {Nature Nanotechnology},
month = {apr},
number = {4},
pages = {421--425},
title = {{Room-temperature nonlinear Hall effect and wireless radiofrequency rectification in Weyl semimetal TaIrTe4}},
url = {https://www.nature.com/articles/s41565-020-00839-3},
volume = {16},
year = {2021}
}

@article{SHC_topo_2023,
  title = {Enhanced spin Hall response from aligned Kramers-Weyl points in high Chern number semimetals},
  author = {Ascencio, C. O. and Jiang, Wei and de Sousa, D. J. P. and Lee, Seungjun and Wang, Jian-Ping and Low, Tony},
  journal = {Phys. Rev. B},
  volume = {108},
  issue = {20},
  pages = {L201404},
  numpages = {6},
  year = {2023},
  month = {Nov},
  publisher = {American Physical Society},
  doi = {10.1103/PhysRevB.108.L201404},
  url = {https://link.aps.org/doi/10.1103/PhysRevB.108.L201404}
}

@article{Hsieh2022,
  title = {Helicity-tunable spin Hall and spin Nernst effects in unconventional chiral fermion semimetals $XY$ ($X=\mathrm{Co},\mathrm{Rh}$; $Y=\mathrm{Si},\mathrm{Ge}$)},
  author = {Hsieh, Ting-Yun and Prasad, Babu Baijnath and Guo, Guang-Yu},
  journal = {Phys. Rev. B},
  volume = {106},
  issue = {16},
  pages = {165102},
  numpages = {12},
  year = {2022},
  month = {Oct},
  publisher = {American Physical Society},
  doi = {10.1103/PhysRevB.106.165102},
  url = {https://link.aps.org/doi/10.1103/PhysRevB.106.165102}
}

@article{Zhang2021SHC,
author = {Zhang, Yang and Xu, Qiunan and Koepernik, Klaus and Rezaev, Roman and Janson, Oleg and {\v{Z}}elezn{\'{y}}, Jakub and Jungwirth, Tom{\'{a}}{\v{s}} and Felser, Claudia and van den Brink, Jeroen and Sun, Yan},
doi = {10.1038/s41524-021-00635-0},
issn = {2057-3960},
journal = {npj Computational Materials},
month = {oct},
number = {1},
pages = {167},
title = {{Different types of spin currents in the comprehensive materials database of nonmagnetic spin Hall effect}},
url = {https://www.nature.com/articles/s41524-021-00635-0},
volume = {7},
year = {2021}
}

@article{Li2026_topo+super,
  title = {{Topological transition and superconducting dome in ${\mathrm{W}}_{2}\mathrm{C}$ tuned by nitrogen doping}},
  author = {Li, Ya-Ping and Qiao, Shu-Xiang and Ni, Mei-Yan and Lu, Hong-Yan and Zhang, Ping},
  journal = {Phys. Rev. B},
  volume = {114},
  issue = {27},
  pages = {274506},
  numpages = {8},
  year = {2026},
  month = {Nov},
  publisher = {American Physical Society},
  doi = {10.1103/ngm6-z5xx},
  url = {https://link.aps.org/doi/10.1103/ngm6-z5xx}
}

@book{Bradley2009,
  author    = {Bradley, C. J. and Cracknell, A. P.},
  title     = {The Mathematical Theory of Symmetry in Solids:
               Representation Theory for Point Groups and Space Groups},
  publisher = {Oxford University Press},
  series    = {Oxford Classic Texts in the Physical Sciences},
  year      = {2009},
  isbn      = {9780191576898}
}

@article{Elcoro2017,
author = "Elcoro, Luis and Bradlyn, Barry and Wang, Zhijun and Vergniory, Maia G. and Cano, Jennifer and Felser, Claudia and Bernevig, B. Andrei and Orobengoa, Danel and de la Flor, Gemma and Aroyo, Mois I.",
title = "{Double crystallographic groups and their representations on the Bilbao Crystallographic Server}",
journal = "Journal of Applied Crystallography",
year = "2017",
volume = "50",
number = "5",
pages = "1457--1477",
month = "Oct",
doi = {10.1107/S1600576717011712},
url = {https://doi.org/10.1107/S1600576717011712},
}

@article{Gracia2021, 
  title = {Anisotropic phonon-mediated electronic transport in chiral Weyl semimetals},
  author = {Garcia, Christina A. C. and Nenno, Dennis M. and Varnavides, Georgios and Narang, Prineha},
  journal = {Phys. Rev. Mater.},
  volume = {5},
  issue = {9},
  pages = {L091202},
  numpages = {6},
  year = {2021},
  month = {Sep},
  publisher = {American Physical Society},
  doi = {10.1103/PhysRevMaterials.5.L091202},
  url = {https://link.aps.org/doi/10.1103/PhysRevMaterials.5.L091202}
}

@article{Ulrich2026,
  title = {Quantum geometric origin of the intrinsic nonlinear Hall effect},
  author = {Ulrich, Yannis and Mitscherling, Johannes and Classen, Laura and Schnyder, Andreas P.},
  journal = {Phys. Rev. B},
  volume = {113},
  issue = {20},
  pages = {L201107},
  numpages = {9},
  year = {2026},
  month = {May},
  publisher = {American Physical Society},
  doi = {10.1103/4z8z-4kch},
  url = {https://link.aps.org/doi/10.1103/4z8z-4kch}
}

@article{Palvi2023,
  title = {Essay: Where Can Quantum Geometry Lead Us?},
  author = {T{\"o}rm{\"a}, P{\"a}ivi},
  journal = {Phys. Rev. Lett.},
  volume = {131},
  issue = {24},
  pages = {240001},
  numpages = {7},
  year = {2023},
  month = {Dec},
  publisher = {American Physical Society},
  doi = {10.1103/PhysRevLett.131.240001},
  url = {https://link.aps.org/doi/10.1103/PhysRevLett.131.240001}
}

@article{Yao2025,
  title = {Observation of chiral surface state in superconducting ${\mathrm{NbGe}}_{2}$},
  author = {Yao, Mengyu and Gutierrez-Amigo, Martin and Roychowdhury, Subhajit and Errea, Ion and Fedorov, Alexander and Strocov, Vladimir N. and Vergniory, Maia G. and Felser, Claudia},
  journal = {Phys. Rev. Mater.},
  volume = {9},
  issue = {3},
  pages = {034803},
  numpages = {7},
  year = {2025},
  month = {Mar},
  publisher = {American Physical Society},
  doi = {10.1103/PhysRevMaterials.9.034803},
  url = {https://link.aps.org/doi/10.1103/PhysRevMaterials.9.034803}
}

\end{document}


	
	\title{\textbf{Supplemental Material \\
		for \\
		``Multicharged Kramers--Weyl Fermions, Quantum Geometric Hall Responses and de Haas--van Alphen Oscillations in WAl$_2$"} 
}


	
	\author{Saurav Patel}
	\affiliation{Department of Physics, Faculty of Science, The Maharaja Sayajirao University of Baroda, Vadodara-390002, Gujarat, India.}
		
	\author{Meet Shah}
	\affiliation{Department of Physics, Faculty of Science, The Maharaja Sayajirao University of Baroda, Vadodara-390002, Gujarat, India.}
	
	\author{Prafulla K. Jha}
	\email{prafullaj@yahoo.com}  
	\affiliation{Department of Physics, Faculty of Science, The Maharaja Sayajirao University of Baroda, Vadodara-390002, Gujarat, India.}

\maketitle
\clearpage

\textbf{Section I:} \textbf{Thermal dispersion curves of
$\boldsymbol{\mathit{WAl}_2}$ investigated through
ab initio molecular dynamics simulations.}

\begin{figure}[H]
	\centering
	\includegraphics[width=0.85\linewidth]{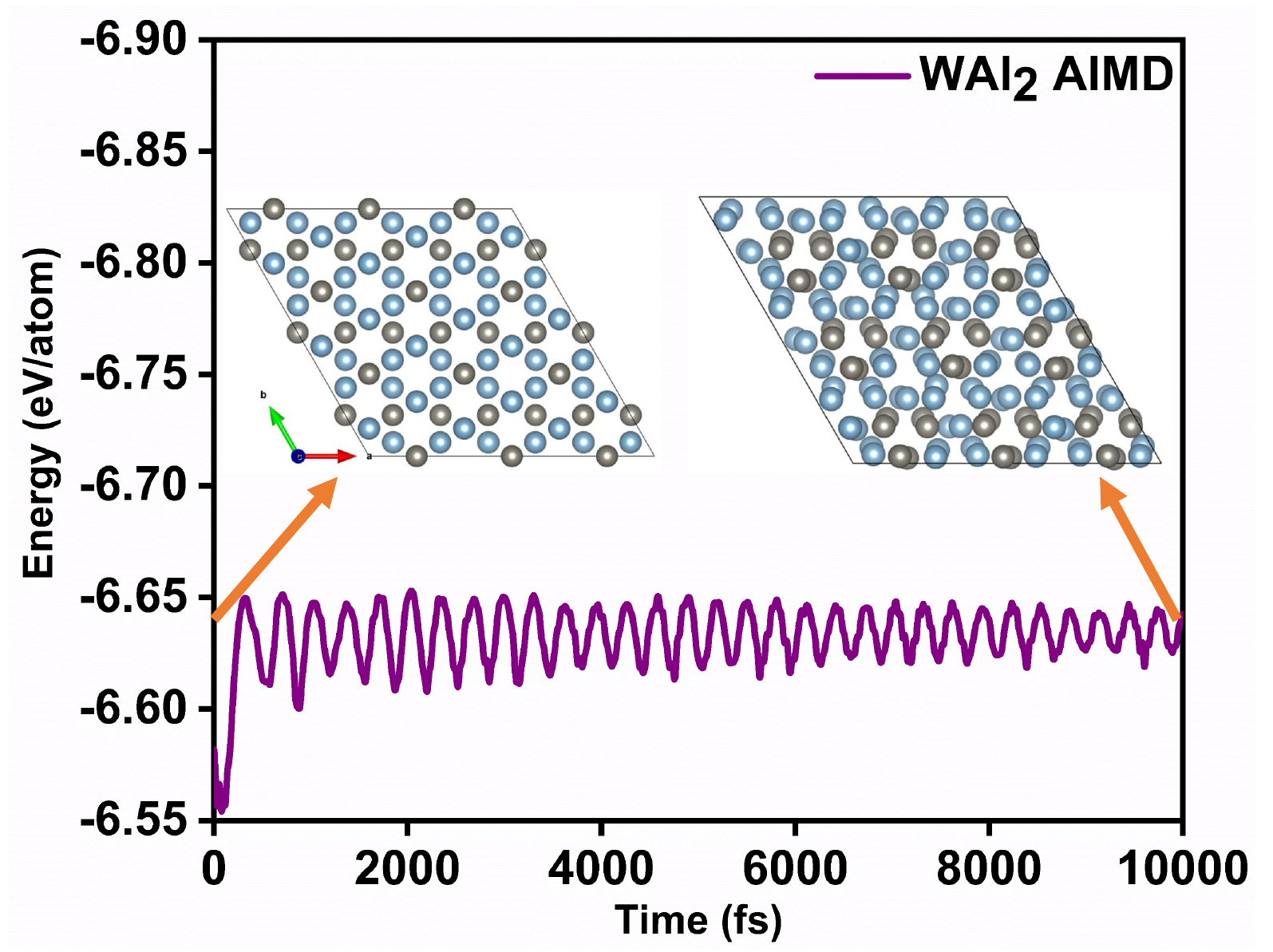}
	\caption{Ab initio molecular dynamics (AIMD) simulations of WAl$_2$ at 300 K}
	\label{Fig:AIMD}
\end{figure}

\clearpage

\textbf{Section II:} \textbf{The orbital projected band stucture of WAl$_2$ }
\begin{figure}[H]
	\centering
	\includegraphics[width=1.0\linewidth]{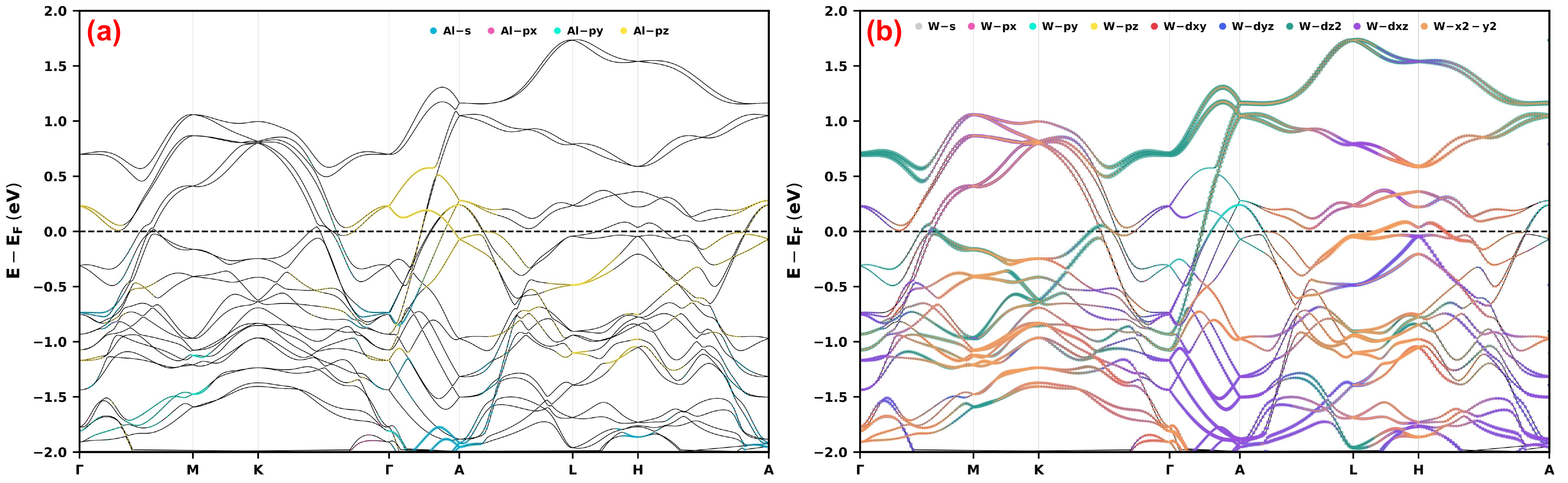}
	\caption{Orbital resolved electronic band structure of WAl$_2$ along the high symmetry paths of the Brillouin zone. (a)Contributions from Al-$s$ and Al-$p$ orbitals ($p_x$, $p_y$, and $p_z$). (b) Contributions from W-$s$, W-$p$, and W-$d$ orbitals, including the five $d$-orbital components.}
	\label{Fig:orbital}
\end{figure}

\clearpage

\textbf{Section III:} \textbf{Comparison of DFT and Wannier interpolated band stucture of WAl$_2$}
\begin{figure}[H]
	\centering
	\includegraphics[width=1.0\linewidth]{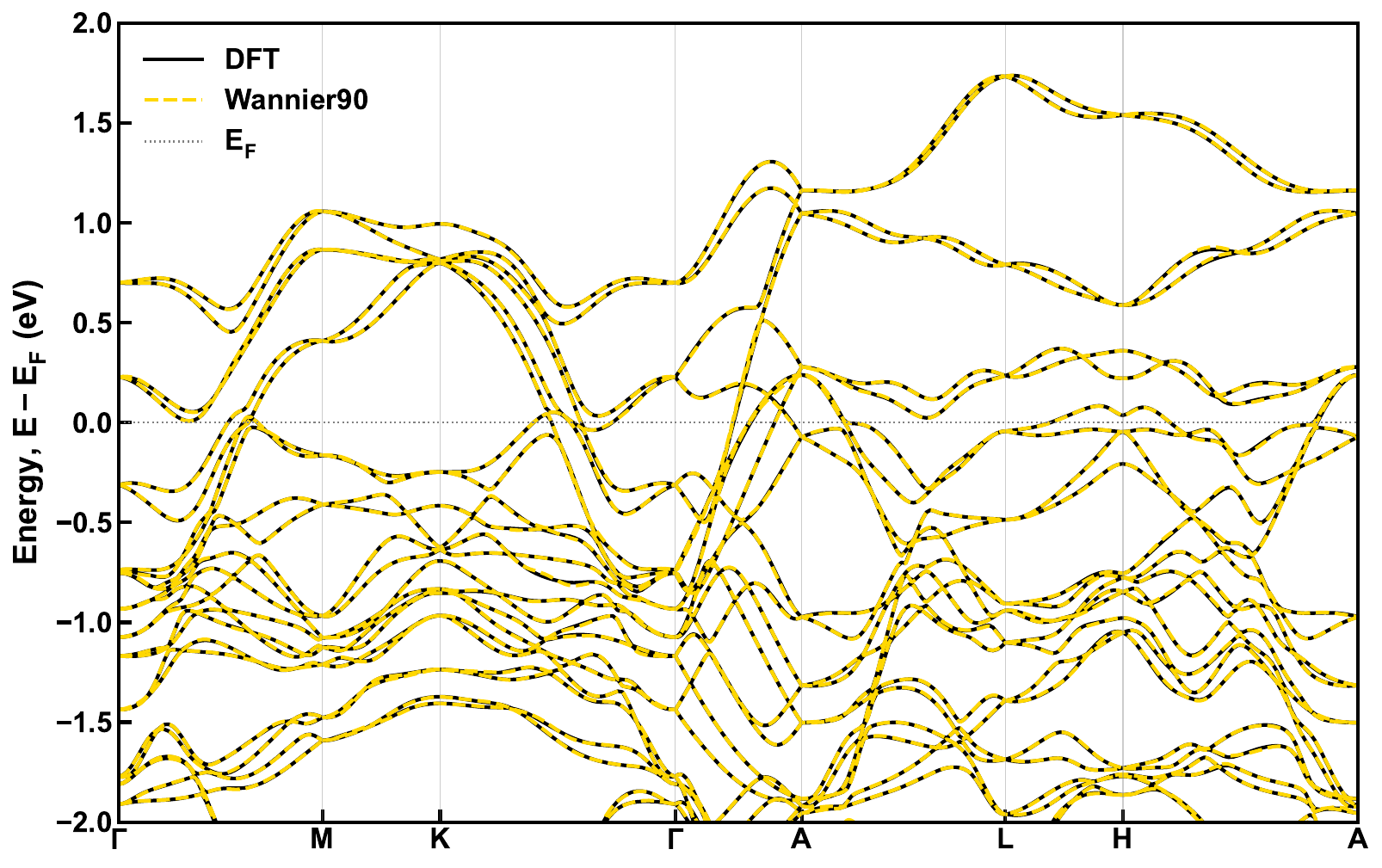}
	\caption{Electronic band structure of WAl$_2$, comparing DFT (solid black) with the Wannier90 interpolated bands (dashed gold) within the energy window of -2 to 2 eV.}
	\label{Fig:W90}
\end{figure}

\clearpage
\textbf{Section IV:} \textbf{Band Structure along with Density of States}
\begin{figure}[H]
	\centering
	\includegraphics[width=1.0\linewidth]{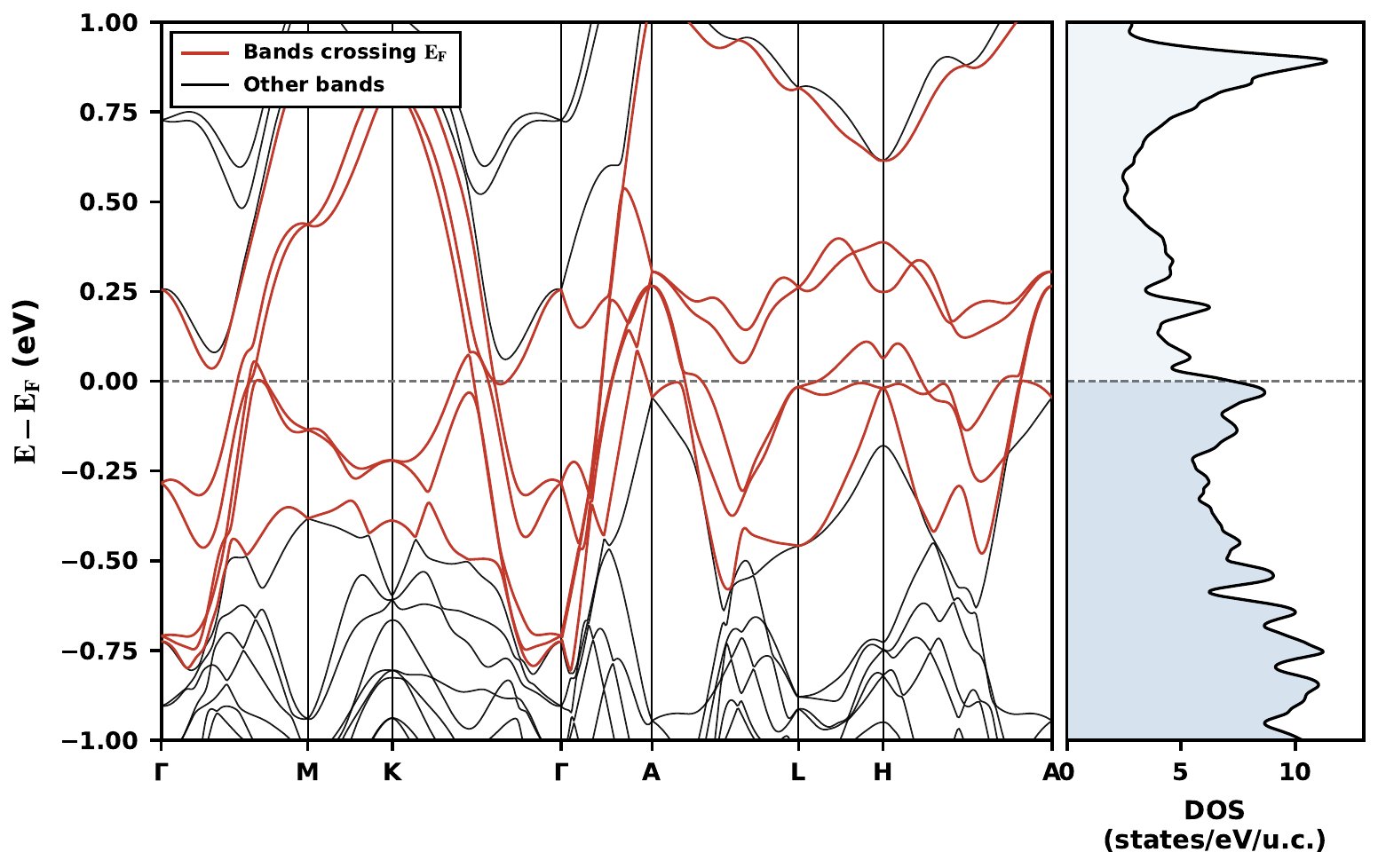}
	\caption{Relativistic band structure of WAl$_2$ (left) and total density of states (right), with the bands crossing the Fermi energy highlighted in red}
	\label{DOS}
\end{figure}
\clearpage

\textbf{Section V:} \textbf{de Hass-Van Alphen low frequency splitting}
\begin{figure}[H]
	\centering
	\includegraphics[width=1.0\linewidth]{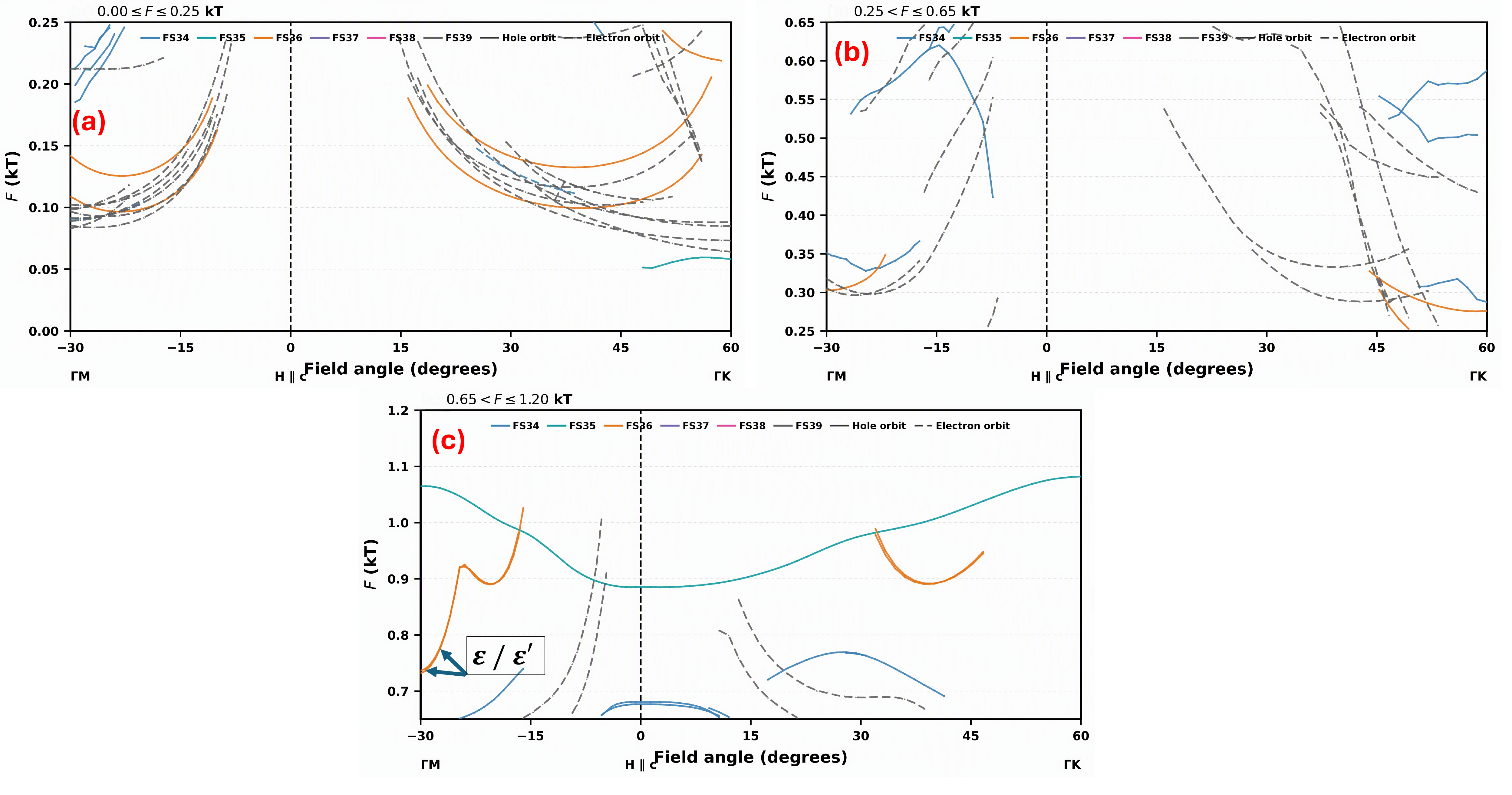}
	\caption{Angular dependence of the calculated de Haas--van Alphen (dHvA) frequencies associated with the Fermi-surface sheets FS34--FS39 of WAl$_2$. The frequency branches are separated into three ranges: (a) $0.00 \leq F \leq 0.25$ kT, (b) $0.25 < F \leq 0.65$ kT, and (c) $0.65 < F \leq 1.20$ kT. The magnetic field is rotated within the $\Gamma$--M and $\Gamma$--K planes, with $H\parallel c$ corresponding to zero field angle. Different colours distinguish the individual Fermi-surface sheets, while solid and dashed curves represent hole-like and electron-like extremal orbits, respectively. The $\epsilon/\epsilon'$ pair in panel (c) indicates a candidate ASOC-induced splitting for the field orientation along the $\Gamma$--M direction.}
	\label{low_spliting}
\end{figure}

\clearpage
\textbf{Section VI:} \textbf{Symmetry analysis of the Kramers--Weyl nodes}

We investigated the symmetry protection of the Kramers--Weyl (KW) nodes identified at the four time reversal invariant momenta (TRIMs), $\Gamma$, $A$, $M$ and $L$, of WAl$_2$. In the presence of spin--orbit coupling (SOC), time reversal symmetry satisfies $\mathcal{T}^2=-1$. Since all four momenta are invariant under time reversal, i.e., $-\mathbf{k}=\mathbf{k}+\mathbf{G}$, Kramers' theorem enforces a twofold degeneracy at each of these points. The corresponding unitary little co- groups have point group type $622$ ($D_6$), generated by $C_6^+$ and $C_2'$, at $\Gamma$ and $A$, and $222$ ($D_2$), generated by $C_2$ and $C_2''$, at $M$ and $L$.

The individual crossings obtained in our band structure are classified according to the double valued spinor representations of their respective little groups. The corresponding rotation matrices and eigenvalues are summarized in Table~II of the main manuscript. At $\Gamma$, the three identified Kramers crossings belong to the representations $R_7$, $R_8$ and $R_9$. The $R_7$ doublet has $C_6^+$ eigenvalues $\pm i$, corresponding to a triply charged KWF with $|C|=3$, whereas the $R_8$ and $R_9$ doublets have eigenvalue pairs $e^{\pm i\pi/6}$ and $e^{\pm i5\pi/6}$ which carry $|C|=1$.

At $A$, the little co-group has the same point group type as at $\Gamma$, but the sixfold operation is nonsymmorphic. Specifically, it is a screw rotation $\widetilde{C}_6^+=\{C_6^+|00\tau\}$, with $\tau=\tfrac13$ for SG~180 and $\tfrac23$ for SG~181. The fractional translation introduces an additional momentum-dependent Bloch phase in the representation of the screw operation. In Table~II, this common phase is factored out, and the listed matrices describe the rotational part of the screw operation. The resulting symmetry classification identifies the triply charged crossing with the $R_7$ representation, while the singly charged crossing belongs to the $R_8$ or $R_9$ representation. At $M$ and $L$, the relevant little co-group is $222$ ($D_2$), whose spinor representation is denoted by $R_5$. All the crossings identified at these two points belong to this representation and carry $|C|=1$. Multiple crossing can occur at the same TRIM, where same representation can occur at different energies.

The symmetry analysis validates the allowed magnitudes of the topological charges and the representation based assignments are summarized in Table~II.
